\documentclass[9pt,shortpaper,twoside,web]{ieeecolor}
\usepackage{generic}
\usepackage{cite}
\usepackage{amsmath,amssymb,amsfonts}
\usepackage{algorithmic}
\usepackage{graphics}
\usepackage{textcomp}
\usepackage{amsmath,amssymb,amsfonts}
\usepackage{cite}
\usepackage{graphicx}
\usepackage{textcomp}
\usepackage{epstopdf}
\usepackage[font=footnotesize]{caption}
\usepackage{psfrag}
\usepackage{subfigure}
\usepackage{float}
\usepackage{ntheorem}

\newtheorem{lemma}{Lemma}

\newtheorem{thm}{Theorem}
\newtheorem{prop}{Proposition}

\newtheorem{assum}{Assumption}
\newtheorem{defn}{Definition}

\newtheorem{rem}{Remark}

\def\BibTeX{{\rm B\kern-.05em{\sc i\kern-.025em b}\kern-.08em
    T\kern-.1667em\lower.7ex\hbox{E}\kern-.125emX}}
\makeatletter
\def\fnum@figure{\textcolor{subsectioncolor}{\sf Fig.~\thefigure}}
\def\fnum@table{\textcolor{subsectioncolor}{\sf TABLE~\thetable}}
\makeatother
\begin{document}
\title{Stability Analysis of Nash Equilibrium for 2-Agent Loss-Aversion-Based Noncooperative Switched Systems}
\author{Yuyue~Yan,~\IEEEmembership{Student Member,~IEEE,}
        Tomohisa~Hayakawa,~\IEEEmembership{Member,~IEEE}
%\thanks{The preliminary version of the material in this paper has been presented on the 58th IEEE Conference on Decision and Control, Nice, France.}
\thanks{Yuyue Yan and Tomohisa Hayakawa are with the Department of Systems and Control Engineering, Tokyo Institute of Technology, Meguro, Tokyo 152-8552 Japan.  E-mail: {yan.y.ac@m.titech.ac.jp}, {hayakawa@sc.e.titech.ac.jp.} The first author would like to thank the financial support from Chinese Scholarship Council (CSC).}% <-this % stops a space
%\thanks{Manuscript received March 6, 2020; revised August 26, 2015.}
}

\maketitle

\begin{abstract}
The stability property of the loss-aversion-based noncooperative
switched systems with quadratic payoffs is investigated.
In this system, each agent adopts the lower sensitivity parameter in the myopic pseudo-gradient dynamics for the case of losing utility than gaining utility, and
 both system dynamics and switching events (conditions) are depending on agents' payoff functions.
Sufficient conditions under which agents' state converges towards the Nash equilibrium are derived in accordance with the location of the Nash equilibrium.
In the analysis, the mode transition sequence and interesting phenomena which we call flash switching  are characterized.
We present several numerical examples to illustrate the properties of our results.
\end{abstract}

\vspace{-6pt}
\section{Introduction}
\IEEEPARstart{G}{ame} theory is one of the disciplines concerning the relations between human decision making and resulting phenomena as a whole {(\cite{bauso2016game,hespanha2017noncooperative} and the references therein)}. %, and has been extensively used in studying multi-agent systems.
In noncooperative~systems, each agent is presumed to be fully rational and selfish, and hence aims to increase
its own payoff by adjusting its individual state (strategy).
Under this presupposition,
the agents' selfish dynamic decision behaviors  are typically modeled by the
%best-response dynamics for discrete-time systems \cite{cortes2015self,Cao2018bestresponse} and the
 pseudo-gradient dynamics for continuous-time systems \cite{Rosen1965Existence,Shamma2005Dynamic,dindovs2006better}.
{In such setup}, agents' decision depends on the projection of the agents' payoff
functions onto their own state proportioned by their own sensitivity parameters without having foresight.
%Many works are based on the myopic pseudo-gradient dynamics.
On the one hand, some incentive mechanisms are proposed with the assumption that agents are following the
pseudo-gradient dynamics, e.g.,
pricing mechanisms \cite{Alpcan2009A} and zero-sum tax/subsidy approach \cite{yan2019Bifurcation,yan2019Social}.
On the other hand, some issues in pseudo-gradient dynamics are discussed for different scenarios.~ {For example, the impact of quantized communication \cite{nekouei2016performance}, leader-following consensus \cite{Ye2017Distributed}, augmented gradient-play dynamics \cite{Gadjov2017APassivity},
external disturbance \cite{romano2018dynamic}, and redistributive
side payments \cite{Hurst2017} were investigated.} %The applications of game theoretic approach inspired by the pseudo-gradient dynamics are found in numerous fields, e.g., communication networks \cite{Jiang2014,vassaki2016state}, incentive schemes \cite{su2016game}, pricing mechanisms \cite{ye2016game,zou2019joint}, to name but a few.%approach for the scenario where the entire state is not available for all the agents since communication graph is incomplete .

However, %as the important part of behavioral economics,
psychological game theory shows by experimental research that it is inaccurate to simply assume
that all the agents are fully rational and selfish because the agents may have
some social and psychological considerations such as the influence of fairness, guilt aversion, hesitation, and inequality
aversion in the decision making \cite{Azar2019}.
On the basis of various psychological considerations, agents make their decisions in  significantly different ways \cite{DanielProspect,pang2017two}.
%For example, it follows from the prospect theory that agents' decision behaviors can be affected by the risks.
%In particular, a risk-averse agent in the noncooperative systems with stochastic payoff functions prefers the option with a lower but guaranteed payoff rather than the option of earning a high payoff along with a chance of losing payoffs, whereas a risk-loving agent prefers the opposite option \cite{pang2017two}.
Therefore, the  pseudo-gradient dynamics with static sensitivity by ignoring all psychological considerations seem unnatural
to describe agents' behavior in the real society.
To our knowledge, the paper \cite{bowling2001convergence} is the first work to characterize the pseudo-gradient dynamics with
variable sensitivity.
The essence of \cite{bowling2001convergence} is to consider the situation where each agent makes a decision quickly when losing and cautiously when winning in a two-agent iterated
matrix game. %with mixed-strategy equilibrium.
However, for describing agents' different decision making when they are facing losses and gains, \emph{loss-aversion} in cognitive psychology and decision theory \cite{tversky1992advances} tells the completely opposite scenario, that is, agents' decision is more \emph{cautious} when they are expected to lose utilities.
In light of the difference between \emph{loss-aversion} phenomena and psychological consideration in \cite{bowling2001convergence}, it is significant to consider the pseudo-gradient dynamics under the loss-aversion scenario.

%To our knowledge, the paper \cite{bowling2001convergence} is the first work to characterize the pseudo-gradient dynamics with
%variable sensitivity.
%The essence of \cite{bowling2001convergence} is to consider the situation where each agent makes the decision quickly when losing and cautiously when winning in a two-agent %, two-action,
%iterated
%matrix game. % with mixed-strategy equilibrium.
%However, for describing agents' different decision making when they are facing losses and gains, \emph{loss-aversion} in cognitive psychology and decision theory \cite{tversky1992advances} tells the completely opposite scenario, that is, agents' decision is more \emph{cautious} when they are expected to lose utilities.
%In light of the difference between \emph{loss-aversion} phenomena and psychological consideration in \cite{bowling2001convergence}, it is necessary to consider the pseudo-gradient dynamics under the loss-averse scenario.

%Since the noncooperative dynamical systems with variable decision speed are captured as hybrid or switched ones and since the pseudo-gradient dynamics driven by quadratic payoff functions are written as linear systems, we make the literature review for the switched systems and piecewise linear systems in this paragraph.

In the last decade, switched systems, which are characterized by a signal specifying the mode transition among a finite number of subsystems, have widely applied to numerous areas. % \cite{kusters2018switch,protasov2019comprehensive,mu2020stability}
As the most important issues in control systems, stability properties of an equilibrium in   switched systems have been extensively investigated \cite{zhao2011stability,zong2016finite,li2018stability,karabacak2019almost}.
In terms of piecewise linear systems, Iwatani and Hara characterized the stability problem based on poles and zeros of the subsystems \cite{Iwatani2006Stability}.
%Nishiyama and Hayakawa provided a series of sufficient conditions to determine stability for 2-dimensional switched linear systems and piecewise nonlinear homogeneous systems \cite{Nishiyama2008Optimal,Nishiyama2009ACC}.
An integration approach based on normalized growth rates was formulated as a tool for judging whether the~trajectory is coming closer to the equilibrium or not in \cite{Nishiyama2008Optimal}.
In the above works, the triggers of mode transitions  are usually understood as event-driven but the events are assumed to be independent of the systems' dynamics.
The fundamental problems on stability and switching behaviors for the special class of switched systems with correlative dynamics and switching events (conditions) get few attentions.

In this paper, we focus on the stability problem for 2-agent noncooperative switched systems, which are characterized as \emph{payoff-driven} piecewise linear systems for describing agents' dynamic decision making with the quadratic payoffs and loss-aversion phenomena. Specifically, we assume that each agent adopts lower sensitivity in the pseudo-gradient dynamics for the case of losing utility than gaining utility
and hence
both the system dynamics and the switching instants depend on the agents' payoff functions.
To determine stability property, %of the Nash equilibrium, %loss-aversion-based noncooperative switched systems,
we characterize the domains in which agents' payoffs are either increasing or decreasing. %, and use the normalized radial growth rate for the Nash equilibrium.
%By assuming that the agents keep on rotating, we reveal an interesting property of agents' decision behaviors in terms of the consistent rotational direction of the trajectories in the state space.
%The  is found in \cite{yan2019aversion}.
Different from preliminary version \cite{yan2019aversion}, this paper categorizes the loss-aversion-based noncooperative systems to 3 cases in accordance with the location of the Nash equilibrium relative to the 2 payoff functions and comprehensively analyzes the differences between the 3 cases. % in terms of mode transition. %and normalized radial growth rates.
%This work contributes to the piecewise linear systems in a payoff-driven fashion.
Observing the fact that the Nash equilibrium is  on the boundaries of the aforementioned domains, by making the approximation for the domains around the Nash equilibrium, we characterize the partition of the state space and the mode transitions to form a piecewise linear (linearized) system.
Moreover, we observe an interesting phenomenon that we call a flash switching  where a single agents' sensitivity transition makes
the other agent immediately switch its sensitivity almost at the
same time instant, and we characterize the necessary condition for a switching instant holding such a phenomenon.

The paper is organized as follows. We formulate the noncooperative system with  the loss-aversion-based pseudo-gradient dynamics
in Section~\ref{formulation}.  {In Section~\ref{partition}, we characterize the domains in which agents keep gaining/losing utilities for 3 cases depending on the relations between the agents' payoff functions.} In Section~\ref{results}, we investigate  stability properties of the Nash equilibrium for the 3 cases  by integration of the normalized radial growth rates.
%The general properties and the differences between the 3 cases in terms of agents' decision behavior are further investigated.
We present a couple of illustrative numerical examples in Section~\ref{numerical}.
Finally, we conclude our paper in Section~\ref{conclusion}.

\emph{Notations}. We use a fairly standard notation in the paper. Specifically, we write $\mathbb{R}$ for the set of real numbers, $\mathbb{R}_+$ for positive real numbers, $\mathbb{R}^{n}$ for the set of \emph{n}$\times 1$ real column vectors, $\mathbb{R}$$^{n \times m}$ for the set of \emph{n}$\times m$ real matrices, $\wedge$ for the logical conjunction, and $\vee$ for the logical disjunction. Moreover, $(\cdot)^\mathrm{T}$ denotes transpose, $\det(\cdot)$ denotes the determinant, and
${\rm diag}[\cdot]$ denotes a diagonal matrix.

\vspace{-10pt}
\section{Problem Formulation}\label{formulation}\vspace{-2pt}
\subsection{Noncooperative Systems with Quadratic Payoffs}\vspace{-3pt}
Consider the noncooperative system with 2 agents selfishly controlling their individual state $x_{i} \in \mathbb{R}$, $i\in\{1,2\}$.
Let $x=[x_1, x_{2}]^{\mathrm T}\in \mathbb{R}^{2}$ denote the agents' state profile.
In this paper, we consider the situation where each agent $i$ aims to increase its own payoff function $J_i(x_i,x_{j})$,
where $J_i:\mathbb{R}^{2}\to \mathbb{R}$ and $j$ is the opponent of agent $i\not=j$. We denote the noncooperative system by $\mathcal{G}(J)$ with $J\triangleq\{J_1,J_2\}$.

%\begin{assum} \label{assum1} \emph{The payoff functions $J_i(x)$, $i\in\{1,2\}$, are twice continuously differentiable and strictly concave with respect to $x_i$, i.e., $\frac{\partial^2 J_i(x)}{\partial x_i^2}<0$, $x\in\mathbb{R}^{2}$, $i\in\{1,2\}$.}\end{assum}
\vspace{-5pt}
\begin{defn} \label{def1} For the noncooperative system $\mathcal{G}(J)$, the state profile $x^{*}\in \mathbb{R}^{2}$ is called a Nash equilibrium of $\mathcal{G}(J)$ if\vspace{-4pt}
\begin{eqnarray}
J_i(x_i^*,x_{j}^*)\geq J_i(x_i,x_{j}^*),\quad  x_i\in\mathbb{R},\quad i\in\{1,2\}.\vspace{-4pt}
\end{eqnarray}
\end{defn}
\vspace{-4pt}

%We use Assumption~\ref{assum1} in all of the following statements.

In this paper, we consider the noncooperative system $\mathcal{G}(J)$ with \emph{quadratic} payoff functions $J_i:\mathbb{R}^{2}\to\mathbb{R}$ given by
{\setlength\abovedisplayskip{1pt}
\setlength\belowdisplayskip{1pt}
 \begin{eqnarray}\label{eqqudractic}
J_i(x)=\frac{1}{2}x^\mathrm{T}A_ix+b_i^\mathrm{T}x+c_i,\quad i\in\{1,2\},\vspace{-4pt}
\end{eqnarray}
where} $A_i\triangleq\left[\begin{array}{ccc}a^i_{11} & a^i_{12}\\ a^i_{12} & a^i_{22}\end{array}\right]\in\mathbb{R}^{2\times2}$ with $a^i_{ii}<0$ {(indicating that $J_i(x)$ is strictly concave with respect to $x_i$)} and $a_{11}^1a_{22}^2\not=a_{12}^1a_{12}^2$, $b_i\triangleq [b_1^i,b_2^i]^\mathrm{T}\in\mathbb{R}^2$, and $c_i\in\mathbb{R}$, $i\in \{1,2\}$.
It is important to note that there exists a \emph{unique} Nash equilibrium $x^*$ in $\mathcal{G}(J)$ in the unbounded state space satisfying
{\setlength\abovedisplayskip{1pt}
\setlength\belowdisplayskip{1pt}
\begin{eqnarray}
\label{approximated2}
0=\frac{\partial J_1(x)}{\partial x_1}=a_{11}^1x_1+a_{12}^1x_2+b_1^1, \\ \label{approximated}
0=\frac{\partial J_2(x)}{\partial x_2}=a_{12}^2x_1+a_{22}^2x_2+b_2^2,\vspace{-4pt}
\end{eqnarray}
 {for $x=x^*$.}
Specifically, the \emph{unique} Nash equilibrium is given by
{\setlength\abovedisplayskip{1.2pt}
\setlength\belowdisplayskip{1.2pt}
\begin{eqnarray}\label{Nash}
x^*=-\left[\begin{array}{ccc}a^1_{11} &a^1_{12}  \\ a^2_{12}  & a^2_{22} \end{array}\right]^{-1}\left[\begin{array}{ccc}b^1_1  \\  b^2_2 \end{array}\right],
\end{eqnarray}
 {because the condition $a_{11}^1a_{22}^2\not=a_{12}^1a_{12}^2$  implies the inverse exists.}
Notice} that the straight lines (\ref{approximated2}) and (\ref{approximated}) are understood as the \emph{best-response lines} for agents $1$ and $2$, respectively.

\vspace{-12pt}
\subsection{Loss-Aversion-Based Myopic Pseudo-Gradient Dynamics}\vspace{-2.6pt}
In this paper, we consider the situation where each agent selfishly and continually changes its state in the noncooperative system $\mathcal{G}(J)$. We
%suppose the communication network between agents is complete, i.e.,
suppose the state profile $x(\cdot)$ is available for both the agents. In addition, associated with agents' payoff functions $J_1,J_2$, the pseudo-gradient dynamics are used to describe agents' myopic selfish behaviors given by {\setlength\abovedisplayskip{1.2pt}
\setlength\belowdisplayskip{1.2pt}
\begin{eqnarray}\label{eqdynamics1}
\dot{x}_i(t)=\alpha_i(t)\frac{ \partial J_i(x(t))}{\partial x_i}, \quad i\in\{1,2\},
\end{eqnarray} where} $\alpha_1(t)$, $\alpha_2(t)\in\mathbb R_+$ are agents' personal (private) sensitivity parameters.
 {Note that the pseudo-gradient dynamics capture the fact that the agents  concern their own payoffs and myopically change their states according to the current information without any foresight on the future state.}

Different from the models in \cite{Ye2017Distributed,Gadjov2017APassivity}, where $\alpha_1(t),\alpha_2(t)$ are constant, in this paper we suppose that {each agent directly observes its own payoff level $J_i(t)$ for agent $i$, i.e., the payoff level $J_i$ is \emph{not calculated} by (\ref{eqqudractic}) through the knowledge of $x$. As such, each agent is supposed to be able to evaluate $\dot J_i$ at infinitesimally small previous time instant $t^-$. Furthermore,} the agents' sensitivity parameters $\alpha_1,\alpha_2$ are \emph{piecewise constant} between 2 values following the loss-aversion-based psychological consideration defined by {\setlength\abovedisplayskip{1.2pt}
\setlength\belowdisplayskip{1.2pt}
\begin{eqnarray}\label{eqdynamics2}
\alpha_i(t)\triangleq\left\{\begin{array}{c}
                        \alpha_i^{\rm L},\quad \quad  {{\rm if\ }\dot J_i(t^-)<0},\\
                        \alpha_i^{\rm H},\quad \quad  {{\rm if\ }\dot J_i(t^-)>0},
    %                    \tilde\alpha_i(t^-), \quad \!\!{\rm if\ }\dot J_i(x(t^-))=0,
                      \end{array}
\right. \quad i\in\{1,2\},
\end{eqnarray}
where} $\alpha_i^{\rm L},\alpha_i^{\rm H}\in\mathbb R_+$ capture the sensitivity of the change of agent $i$'s state per unit time against losing and gaining payoff environment, respectively, for $i\in\{1,2\}$.
%$\tilde\alpha_i(t)\triangleq\{\alpha_i^{\rm L},\alpha_i^{\rm H}\}\setminus\alpha_i(t)$ denotes the opposite sensitivity parameter instead of the current one at time $t$.
 {As soon as agent $i$ reaches the state observing $\dot J_i(t)=0$, it switches its $\alpha_i(t)$ to the other value.}
 {An interesting observation that this sensitivity parameter change by one agent may give rise to the parameter change of the other agent is elaborated in Section~\ref{section_C} below.}

We connect the phenomenon of \emph{loss-aversion} in \emph{prospect theory} \cite{DanielProspect} with the noncooperative behaviors in $\mathcal G(J)$.
It is well known that humans are more cautious to make the decision when they face losing payoff than gaining payoff.
As a typical example, in the stock investment market, investors (agents) have the tendency to hold losing investments very long and sell winning investments very soon \cite{odean1998investors}.
In light of this observation, in this paper
we suppose that the sensitivity
parameters satisfy $\alpha_i^{\rm L}\leq\alpha_i^{\rm H}$, $i\in\{1,2\}$, to describe agents' slower  behavior for the case where their corresponding {$\dot J_i(t)$} is negative.
\begin{figure*}
\begin{minipage}{0.31\textwidth}
\centering
\includegraphics[width=5.2cm]{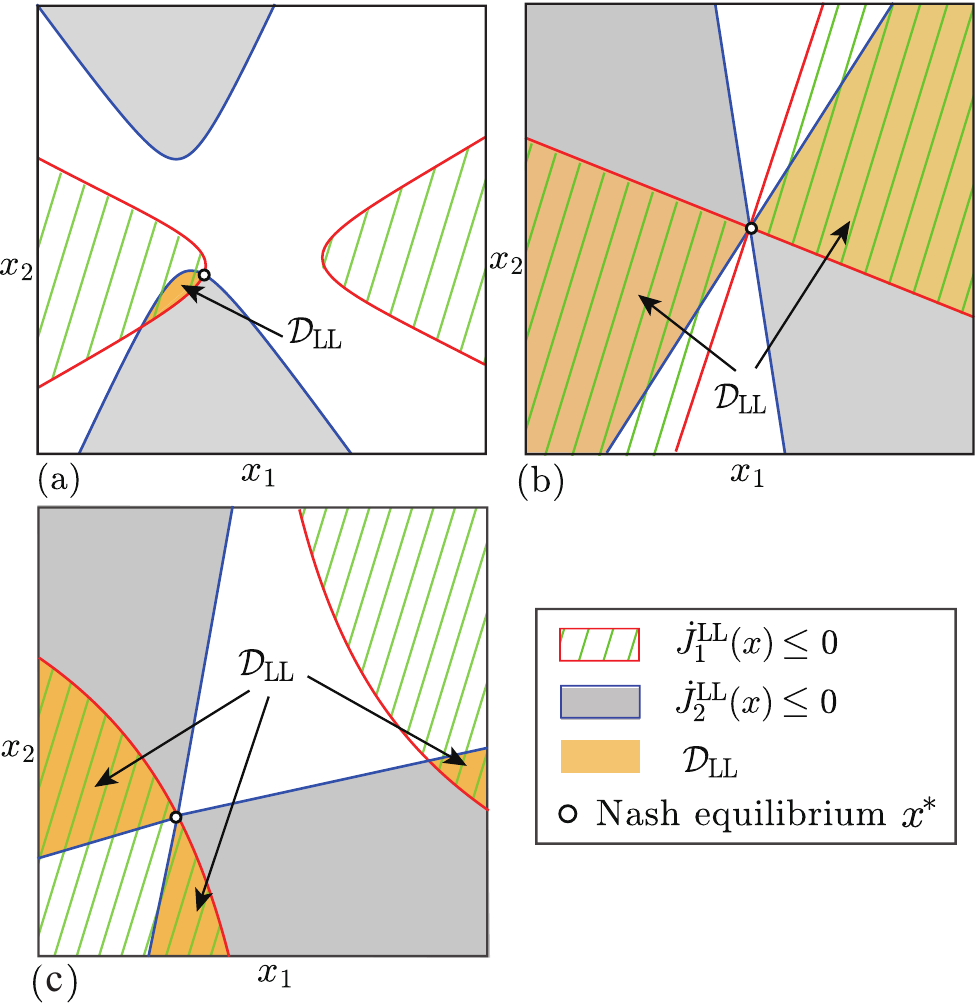}    % The printed column width is 8.4 cm.
\caption{\unboldmath Examples of the domain $\mathcal D_{\rm LL}$.  {(a): Case 1, (b): Case 2, (c): Case 3.}
The other domains $\mathcal D_{\rm HL}$, $\mathcal D_{\rm LH}$, $\mathcal D_{\rm HH}$ can be similarly characterized.
}
		\label{Region_LL}
\end{minipage}\hspace{3pt}
\begin{minipage}{0.30\textwidth}
\centering
	\includegraphics[width=5.2cm]{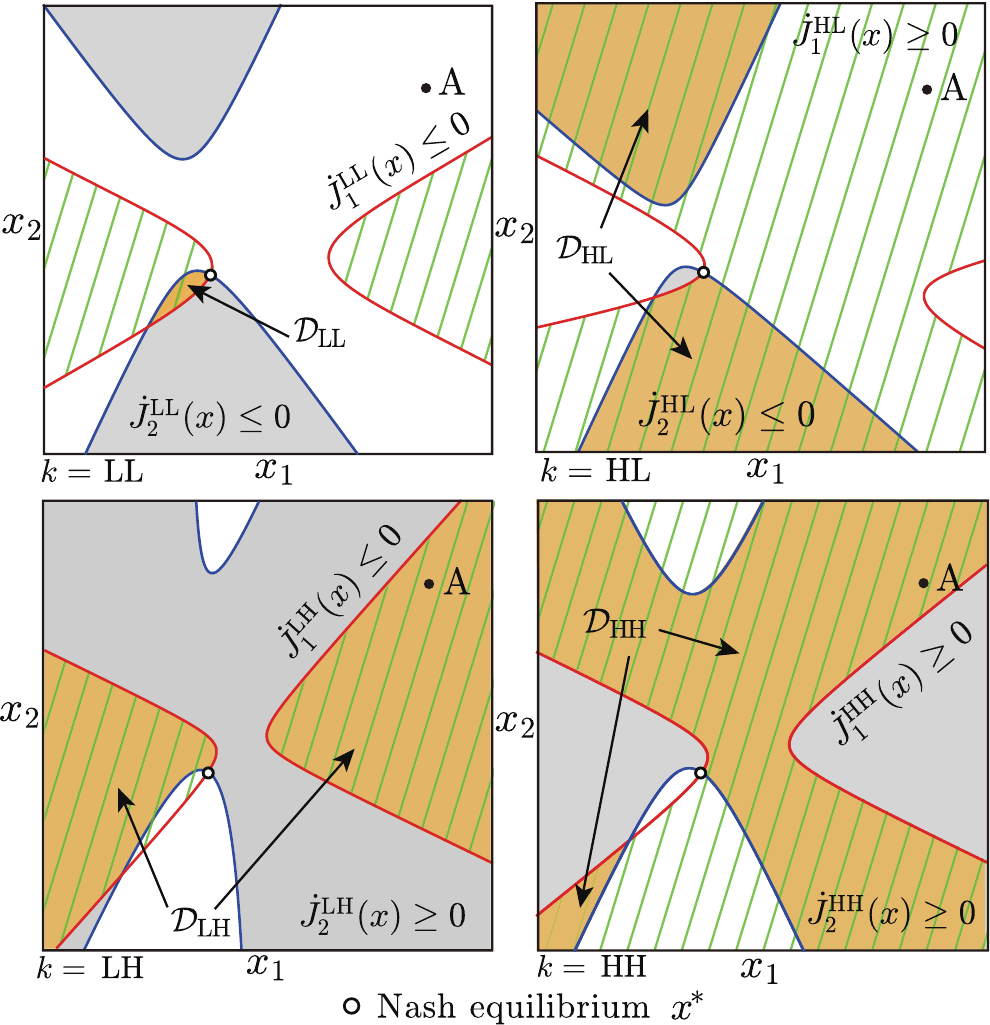}    % The printed column width is 8.4 cm.
\caption{\unboldmath An example of the 4 domains $\mathcal D_{\rm LL}$, $\mathcal D_{\rm HL} $, $\mathcal D_{\rm LH}$, $\mathcal D_{\rm HH}$ {for Case 1}. The figure for the case of $k=\rm LL$ is the copy of Fig.~\ref{Region_LL}(a).}
		\label{Region_Four}
\end{minipage}\hspace{4pt}
\begin{minipage}{0.36\textwidth}
\centering
		\includegraphics[width=6.48cm]{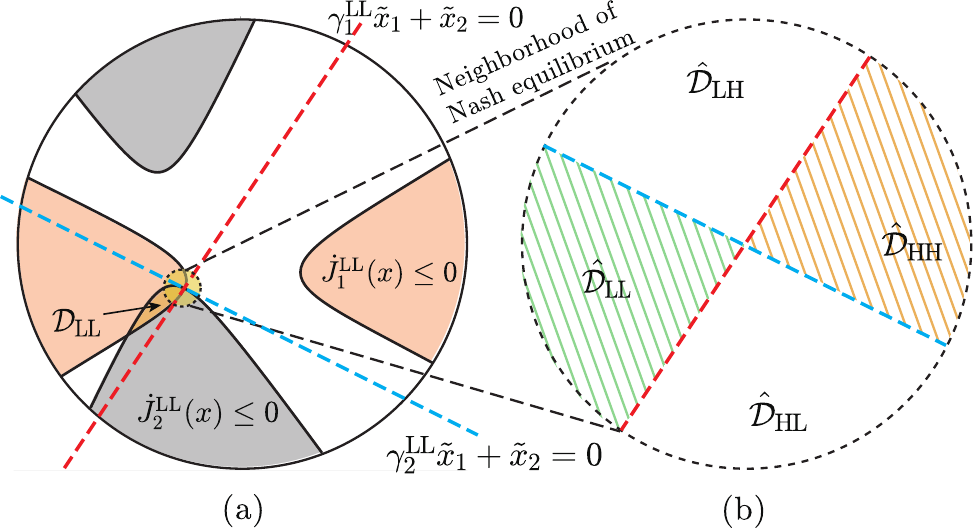}    % The printed column width is 8.4 cm.
		 \caption{\unboldmath Approximated domain where $a_{12}^2(a_{12}^1 x_1^*+a_{22}^1x_2^*+b_2^1)>0$, $a_{11}^1(a_{11}^2 x_1^*+a_{12}^2x_2^*+b_1^2)>0$. (a): ${\mathcal D}_{\rm LL}$, (b): the approximated domains $\hat{\mathcal D}_k$, $k\in\mathcal K$, around the neighborhood of Nash equilibrium. The rotational direction is counterclockwise since $\gamma_1^{\rm LL}<0\wedge\gamma_2^{\rm LL}>0$ implies $a_{12}^1<0\wedge a_{12}^2>0$.}
		\label{Region3}
\end{minipage}
\vspace{-15pt}
\end{figure*}
\vspace{-1pt}

It is important to note that there are 4 possibly different combinations (modes) of agents' sensitivities depending on the signs of $\dot J_1$ and $\dot J_2$. Henceforth, we let
{\setlength\abovedisplayskip{1pt}
\setlength\belowdisplayskip{1pt} \begin{eqnarray}\label{table1}
\alpha^{\rm LL}\triangleq{\rm diag}[\alpha_1^{\rm L},\alpha_2^{\rm L}],\quad%\ \dot J_1(x(t))<0,\ \dot J_2(x(t))<0,\\
\alpha^{\rm HL}\triangleq{\rm diag}[\alpha_1^{\rm H},\alpha_2^{\rm L}],\\%\ \dot J_1(x(t))\geq0,\  \dot J_2(x(t))<0,\\
\alpha^{\rm LH}\triangleq{\rm diag}[\alpha_1^{\rm L},\alpha_2^{\rm H}],\quad%\ \dot J_1(x(t))<0, \ \dot J_2(x(t))\geq0,\\
\alpha^{\rm HH}\triangleq{\rm diag}[\alpha_1^{\rm H},\alpha_2^{\rm H}],\vspace{-4pt}%\ \dot J_1(x(t))\geq0, \ \dot J_2(x(t))\geq0,
%\alpha^k\triangleq\left\{\begin{array}{c}
%                        {\rm diag}[\alpha_1^{\rm L},\alpha_2^{\rm L}],\ \dot J_1(x(t))<0,\ \dot J_2(x(t))<0,\\
%                        {\rm diag}[\alpha_1^{\rm H},\alpha_2^{\rm L}],\ \dot J_1(x(t))\geq0,\  \dot J_2(x(t))<0,\\
%                        {\rm diag}[\alpha_1^{\rm L},\alpha_2^{\rm H}],\ \dot J_1(x(t))<0, \ \dot J_2(x(t))\geq0,\\
%                        {\rm diag}[\alpha_1^{\rm H},\alpha_2^{\rm H}],\ \dot J_1(x(t))\geq0, \ \dot J_2(x(t))\geq0,
%                      \end{array}
%\right.
\end{eqnarray}
to} denote the entire sensitivity profile of the 2 agents. Consequently, agents' decision behaviors (\ref{eqdynamics1}) with the loss-aversion-based sensitivity (\ref{eqdynamics2}) and the quadratic payoff functions (\ref{eqqudractic}) under mode $k\in
\mathcal K\triangleq
\{{\rm LL,HL,LH,HH}\}$ are described as
{\setlength\abovedisplayskip{0.5pt}
\setlength\belowdisplayskip{0.5pt} \begin{align}\label{Dynamics_real}
\nonumber\dot{ x}(t)&=\alpha^{k(t)}\left[\frac{ \partial J_1( x(t))}{\partial  x_1},\frac{ \partial J_2( x(t))}{\partial  x_2}\right]^{\mathrm T}\\
 &=\mathbb A_{k(t)}( x(t)-x^*),
\end{align}
where} $\mathbb A_k\triangleq\alpha^k\left[\!\!\begin{array}{ccc}a^1_{11} &a^1_{12}  \\ a^2_{12}  & a^2_{22} \end{array}\!\!\right]$ denotes the system matrix under mode $k\in
\mathcal K$ and $x^*$ is given by (\ref{Nash}).
%Note that since $a_{11}^1a_{22}^2\not=a_{12}^1a_{12}^2$ is assumed, $\mathbb A_k$ is nonsingular for all $k\in\mathcal K$.
 {As discussed in the following sections, it turns out that which mode is active can be characterized by some domains in the state space.}
% {Hence, to investigate the stability property, we focus on characterizing the partition of the state space and finding a piecewise linearized system of (\ref{eqdynamics1}), (\ref{eqdynamics2}) with the characterized partition in the following sections.}

\vspace{-9pt}
\section{{Hyperbolic/Elliptic Domains} Characterizing Utility Trends}\label{partition}\vspace{-4pt}

 %Specifically, we first present some general properties of the system description under the agents' behavior in (\ref{eqdynamics1}), (\ref{eqdynamics2}), and then we show the stability results for the Nash equilibrium $x^*$.

%\subsection{General Properties}\vspace{-3pt}
 {In this section, we characterize the 4 domains associated with the 4 modes in $\mathcal{K}$ depending on the utility trends (increasing or decreasing) of the 2 players. Specifically, we define the 4  domains in which the signs of $\dot{J_1}$ and $\dot{J_2}$ associated with (\ref{eqqudractic}) remain the same to be positive/negative along the system trajectories of (\ref{eqdynamics1}). With a  slight abuse of notation, let the functions $\dot J^k_i\!:\!\mathbb R^2$ $\to\!\mathbb R$ represent the time rate of change $\dot J_i$  of $J_i$ as a function of the state $x$ for  agent $i\!\in\!\{1,2\}$ with mode $k\in\mathcal K$   given by}
{\setlength\abovedisplayskip{1pt}
\setlength\belowdisplayskip{1pt}
\begin{eqnarray}\nonumber
\dot J^k_i(x)\!\!\!&\triangleq&\!\!\!\left[\frac{ \partial J_i( x)}{\partial  x_1},\frac{ \partial J_i(x)}{\partial  x_2}\right]\mathbb A_k (x-x^*)\\ \nonumber
              &=&\!\!\!\textstyle\frac{1}{2} x^{\mathrm T}Q_i^k x+\bigl(\mathbb A_k^{\mathrm T}b_i-A_i\mathbb A_k{x^*}\bigl)^{\mathrm T}x-b_i^{\mathrm T}\mathbb A_k x^*\\ \label{J_ik}
              &=&\!\!\!\textstyle\frac{1}{2} (x-x^*)^{\mathrm T}Q_i^k (x-x^*)+{\beta_i^k}^{\mathrm T}(x-x^*),
\end{eqnarray}
with
}$Q_i^k\triangleq A_i\mathbb A_k+\mathbb A_k^{\mathrm T}A_i\in\mathbb R^{2\times 2}$ and $\beta_i^k\triangleq \mathbb A_k^{\mathrm T}(A_i{x^*}+b_i)\in\mathbb R^{2}$, $i\in\{1,2\},k\in\mathcal K$.  {The function $\dot J^k_i(x)$ is reminiscent of} the time  rate of change of $J_i(\cdot)$ along the system trajectories given by
$\dot J_i(t)=\frac{ \partial J_i(x(t))}{\partial  x}\dot x(t)$ with mode $k$ being active at state~$x$.
%In other words, given a mode $k\in\mathcal K$, $\{ x\in\mathbb R^2:\dot J^{k}_1( x)<0\}$ and $\{ x\in\mathbb R^2:\dot J^{k}_1( x)\geq0\}$
%Obviously, $\dot J_i(x(t))$ is given by $\dot J^k_i(x(t))$ if and only if the active mode at time $t$ is $k\in\mathcal K$.

We
define the domains $ {\mathcal D}_k$, $k\in\mathcal K$, in which each of the agents  {keeps} either the high sensitivity $\alpha_i^{\rm H}$ or the low sensitivity $\alpha_i^{\rm L}$ as
{\setlength\abovedisplayskip{1pt}
\setlength\belowdisplayskip{1pt}
 \begin{align}\label{Dk_LL}
& {\mathcal D}_{\rm LL}\triangleq\{ x\in\mathbb R^2:\dot J^{\rm LL}_1( x)\leq0,\dot J^{\rm LL}_2(x)\leq0\},\\ \label{Dk_HL}
& {\mathcal D}_{\rm HL}\triangleq\{ x\in\mathbb R^2:\dot J^{\rm HL}_1( x)\geq0,\dot J^{\rm HL}_2( x)\leq0\},\\ \label{Dk_LH}
& {\mathcal D}_{\rm LH}\triangleq\{ x\in\mathbb R^2:\dot J^{\rm LH}_1( x)\leq0,\dot J^{\rm LH}_2( x)\geq0\},\\ \label{Dk_HH}
& {\mathcal D}_{\rm HH}\triangleq\{ x\in\mathbb R^2:\dot J^{\rm HH}_1( x)\geq0,\dot J^{\rm HH}_2( x)\geq0\}.
\end{align}
Note that some of these 4 domains may \emph{not} exist (as explained in Remark~\ref{rem_nonexistence_may} below). Furthermore, the Nash equilibrium $x^*$ belongs to all the existing domains, since $\dot J^k_i(x^*)=0$ for all $i\in\{1,2\}$ and $k\in\mathcal K$.

 {It is important to note that the boundaries of $ {\mathcal D}_k$, $k\in\mathcal K$, may be either straight lines or quadratic curves depending on whether $\beta_i^k$ in (\ref{J_ik}) vanishes or not. Specifically, since $\mathbb A_k$, $k\in\mathcal K$, are~nonsingular,  {$A_ix^*+b_i\not=0$ (resp., $A_ix^*+b_i=0$) if and only if ${\beta_i^k}=\mathbb A_k^{\mathrm T}(A_i{x^*}+b_i)\not=0$ (resp., ${\beta_i^k}=0$), $k\in\mathcal K$, so that the boundaries associated with $\dot J_i^k(x) = 0$ are quadratic (hyperbolic/elliptic) curves (resp., straight lines intersected at $x^*$ when $Q_i^k$ is sign-indefinite)}.
Since the domains $ {\mathcal D}_k$, $k\in\mathcal K$, are characterized by the two equations $\dot J_1^k(x)=0$ and $\dot J_2^k(x)=0$, we categorize 3 cases as shown in Fig.~\ref{Region_LL},}
%Figure~ shows some typical 3 cases that we discuss in the following sections,
that is, $A_ix^*+b_i\not=0$ for $i\in\{1,2\}$ (Case~1); $A_ix^*+b_i=0$ for $i\in\{1,2\}$ (Case~2); and $A_1x^*+b_1\not=0$, $A_2x^*+b_2=0$ (Case~3).
 {
%and hence $x^*$ is located at the intersections of quadratic (hyperbolic and/or elliptic) curves $\dot J_1^k(x) =0$ and $\dot J_2^k(x) =0$, $k \in \mathcal K$.
In any case, $x^*$ is always on the cusp of $\mathcal D_k$ for mode $k$ that exists (except for the domain where $\beta_i^k=0$ and $Q_i^k$ is positive definite (see Remark~\ref{rem_nonexistence_may} for an example)).
Here we note that because $A_ix^*+b_i$ is equal to $\frac{\partial J_i(x^*)}{\partial x}$,     the above 3 cases are  categorized according to whether $x^*$ coincides with the maximum (or saddle) point of $J_i(x)$ for agent $i$ (i.e., $\frac{\partial J_i(x^*)}{\partial x}=0$) or not. }
%, i.e., $\frac{\partial J_i(x^*)}{\partial x}=0$.}
%In such a case, the regions of $\{ x\in\mathbb R^2:\dot J^{k}_i( x)\geq0\}$ and $\{ x\in\mathbb R^2:\dot J^{k}_i( x)\leq0\}$, $k\in\mathcal K$, are convex cones (boundaries are straight lines) with $x^*$ being the center (except for a special case explained in Remark~\ref{rem_nonexistence_may}) without any approximation.

\vspace{-5pt}
\begin{rem}\label{rem_nonexistence_may}\emph{
Some of the domains ${\rm int\,}\mathcal D_{k}$, $k\in\mathcal K$, may not exist. For example, consider Case 2 where there exists $\lambda>0$ such that $J_1(x)=\lambda J_2(x)$. In this case, since
$Q_1^k=\lambda A_2\alpha^k{\rm diag}[\lambda,1]A_2+\lambda A_2{\rm diag}[\lambda,1]\alpha^k A_2>0$ and
$Q_2^k= A_2\alpha^k{\rm diag}[\lambda,1]A_2+A_2{\rm diag}[\lambda,1]\alpha^kA_2>0$ hold for all $k\in\mathcal K$,
{it follows that}
${\rm int\,}\mathcal D_{\rm LL}$, ${\rm int\,}\mathcal D_{\rm HL},{\rm int\,}\mathcal D_{\rm LH}=\emptyset$ and $\mathcal D_{\rm HH}=\mathbb R^2$.
Alternatively, consider the case with the zero-sum payoffs, where $J_1(x)=-J_2(x)$. In this case,  since $\dot J_1^k(x)=-\dot J_2^k(x)$ holds for all $x\in\mathbb R^2$ and $k\in\mathcal K$, we have ${\rm int\,}\mathcal D_{\rm LL}={\rm int\,}\mathcal D_{\rm HH}=\emptyset$ and $\mathcal D_{\rm HL}\bigcup\mathcal D_{\rm LH}=\mathbb R^2$.
%Hence, in that case,
% value of $a_{11}^1a_{22}^2-a_{12}^1a_{12}^2$ (see the stability result shown in \cite{yan2019Bifurcation}).
}
\end{rem}

%\begin{figure}
%	\begin{center}
%		\includegraphics[width=7.4cm]{Region_D_exam0-eps-converted-to.pdf}    % The printed column width is 8.4 cm.
%		\vspace{-8pt}\caption{Examples of the domain $\mathcal D_{\rm LL}$. (a): $A_ix^*+b_i\not=0$, $i\in\{1,2\}$, (b): $A_ix^*+b_i=0$, $i\in\{1,2\}$, (c): $A_1x^*+b_1\not=0$, $A_2x^*+b_2=0$.
%The other domains $\mathcal D_{\rm HL}$, $\mathcal D_{\rm LH}$, $\mathcal D_{\rm HH}$ can be similarly characterized.}
%\vspace{-16pt}
%		\label{Region_LL}
%	\end{center}
%\end{figure}
%
%\begin{figure}
%	\begin{center}
%		\includegraphics[width=7.4cm]{coverage-eps-converted-to.pdf}    % The printed column width is 8.4 cm.
%		\vspace{-8pt}\caption{An example of the four domains $\mathcal D_{\rm LL},\mathcal D_{\rm HL},\mathcal D_{\rm LH},\mathcal D_{\rm HH}$ when $A_ix^*+b_i\not=0$, $i\in\{1,2\}$. The figure for case of $k=\rm LL$ is the copy of Fig.~\ref{Region_LL}(a).}
%		\label{Region_Four}
%\vspace{-22pt}
%	\end{center}
%\end{figure}

\vspace{-10pt}
\begin{rem}\emph{
There may exist some overlaps between $\mathcal D_k$, $k\in\mathcal K$. Figure~\ref{Region_Four} shows a typical example of the 4 domains indicated by the orange regions. Note that point $\rm A$ belongs to  the 2 domains $\mathcal D_{\rm LH}$ and $\mathcal D_{\rm HH}$ but not $\mathcal D_{\rm LL}$ nor $\mathcal D_{\rm HL}$.
In defining the mode of the system dynamics (\ref{Dynamics_real}) in the overlapped regions,
the agents keep mode $k$ at time $t^+$ if $x(t)\in{\rm int\,}\mathcal D_k$, given an active mode $k$ at time $t$.
}
\end{rem}

%The following result shows that $\bigcup_{k\in\mathcal K}{\mathcal D}_k=\mathbb R^2$ and hence there exists at least one mode being active in (\ref{eqdynamics1}), (\ref{eqdynamics2}).
\vspace{-12pt}
\begin{lemma}\label{thm_nonzeno}
Consider the loss-aversion-based
noncooperative system $\mathcal G(J)$ with the pseudo-gradient dynamics (\ref{eqdynamics1}), (\ref{eqdynamics2}). Then, it follows that
 $\bigcup_{k\in\mathcal K}{\mathcal D}_k=\mathbb R^2$ for any $\alpha_i^{\rm H}\geq\alpha_i^{\rm L}$, $i=1,2$.
\end{lemma}
\vspace{-4pt}

\vspace{-5pt}
\begin{rem}\emph{
 {Note that if the agents' loss-averse behavior is characterized by $\alpha_i^{\rm H}<\alpha_i^{\rm L}$, $i=1,2$, then $\bigcup_{k\in\mathcal K}{\mathcal D}_k=\mathbb R^2$ may not hold~even though (\ref{Dk_LL})--(\ref{Dk_HH}) are a complete enumeration of all possible cases.}
}
\end{rem}
\vspace{-5pt}

 {Different from the standard  piecewise linear system with conewise partitions \cite{Iwatani2006Stability,Nishiyama2008Optimal}, the main problem in investigating stability property in this paper is to appropriately deal with the overlaps of the domains (Remark 2)  and non-conewise domains. In the following section,
we %categorize the loss-aversion-based noncooperative system to $3$ cases and
introduce how to appropriately partition the state space {depending on the rotational directions of the system trajectories} and  how to characterize stability according to a piecewise linearized system of (\ref{eqdynamics1}), (\ref{eqdynamics2}) whose state is traveling over  the partitioned domains.}
%\begin{rem}\emph{ Theorem~\ref{thm_nonzeno} implies that the mode at time $t^+$ is well-defined after the switch occur at time $t$.
%In other words, suppose at time $t$ agents are leaving from the domain $ {\mathcal D}_i$, $i\in\{1,\ldots,4\}$,
%then there exist at least one
%$j\in\{1,\ldots,4\}\setminus\{i\}$, such that at time $t^+$ agents enter into the domain ${\mathcal D}_j$ and keep the same mode in the nearest future time. In that case, agents' sensitivity profile is understood as switching from mode $i$ to $j$ at the moment $t$.
%For example, if agents break the condition of $\dot J^{\rm LL}_1(x)<0$ (resp., $\dot J^{\rm LL}_2(x)<0$) but keep $\dot J^{\rm LL}_2(x)<0$ (resp., $\dot J^{\rm LL}_1(x)<0$) at time $t$, then agents' sensitivity profile is switching from mode $1$ to mode $2$ (resp., mode $3$). If agents break both of $\dot J^{\rm LL}_1(x)<0$ and $\dot J^{\rm LL}_2(x)<0$, then the mode is switched from $1$ to $4$.
%}
%\end{rem}

\vspace{-11pt}
\section{Stability Analysis}\label{results}\vspace{-2pt}
In this section, we characterize stability properties of the Nash equilibrium $x^*$ for the loss-aversion-based noncooperative system $\mathcal G(J)$. Specifically, we first present the properties of agents' behavior under (\ref{eqdynamics1}), (\ref{eqdynamics2}) in terms of the rotational direction of the trajectories.
% {which help to understand the mode transitions and  appropriately
%partition the state space in the following categorized 3 cases}.
We let $\tilde x\triangleq x-x^*$ and consider the polar form $(r,\theta)$ of the coordinate $(\tilde x_1,\tilde x_2)$.
%Note that the mode $k\in\mathcal K$ and system matrix $\mathbb A_k$ depend solely on $\theta$ but $r$ in this section.
Note that the rotational direction of the trajectories at phase $\theta$ under mode ${k\in\mathcal K}$ can be determined by the sign of
{\setlength\abovedisplayskip{1pt}
\setlength\belowdisplayskip{1pt}
\begin{align}\nonumber
\dot\theta_k&=\frac{{\rm d}}{{\rm d}t}(\tan^{-1}\frac{\tilde x_2}{\tilde x_1})=\frac{-\dot{\tilde  x}_1\tilde x_2+\tilde x_1\dot {\tilde x}_2}{\tilde x_1^2+\tilde x_2^2}=\frac{1}{r^2}\det\left[\begin{array}{ccc}\tilde x_1 & \dot {\tilde x}_1 \\ \tilde x_2  & \dot {\tilde x}_2 \end{array}\right]\\
&=\det[\eta(\theta),\mathbb A_k\eta(\theta)]
=\eta^{\mathrm T}(\theta)P_k\eta(\theta), \label{direction}
\end{align}}\vspace{-0.1pt}\noindent where $\eta(\theta)=[\cos \theta,\sin\theta]^{\mathrm T}$ and
{\setlength\abovedisplayskip{1pt}
\setlength\belowdisplayskip{1pt}
\begin{eqnarray}\label{PK}
\textstyle P_k\triangleq\left[\begin{array}{ccc}\alpha^k_2a_{12}^2 & \frac{-\alpha_1^ka_{11}^1+\alpha_2^ka_{22}^2}{2} \\ \frac{-\alpha_1^ka_{11}^1+\alpha^k_2a_{22}^2}{2}   & -\alpha^k_1a_{12}^1 \end{array}\right],\quad k\in\mathcal K,
\end{eqnarray}
with}  {$\alpha_1^{\rm XY}\triangleq\alpha_1^{\rm X}$, $\alpha_2^{\rm XY}\triangleq\alpha_2^{\rm Y}$, $\rm X,Y\in\{L,H\}$.}
%Therefore, the rotational direction of trajectories depends solely on $\theta$.
In particular, the~trajectories under mode $k\in\mathcal K$ are moving in the counterclockwise (resp., clockwise) direction when $\dot\theta_k>0$ (resp., $\dot\theta_k<0$).

%\begin{rem}\emph{The active mode $k\in\mathcal K$ may not depend only on the phase $\theta$ but also the distance $r$. For example, if $x^*\not=-A_i^{-1}b_i$ holds for some $i\in\{1,2\}$, then $r$ may change the active mode since $\mathcal D_k$, $k\in\mathcal K$, are not cones
% by noting $\dot J^k_i(x)=0$, $k\in\mathcal K$, are not straight lines
%(see Fig.~\ref{Region_Four}).}
%\end{rem}

% {It is well known that} stability analysis for the case with finite number of mode transitions can be made by investigating the stability property of the possible final modes,  {but the one for infinite number of mode transitions is more difficult.
%Hence, we focus on the latter case in the rest of this paper.}
 {To focus on the case where there exist infinitely many mode
transitions for the agents, we assume that the eigenvalues of $\mathbb A_k$ are all complex conjugate in our stability analysis.} The case where there are finite number of mode transitions can be handled by simply investigating the stability property of the possible final modes.

\vspace{-4pt}
\begin{assum} \label{assum4} \emph{The system matrix $\mathbb A_k$ has a pair of complex conjugate eigenvalues for all the modes $k\in\mathcal K$.}\end{assum}
\vspace{-4pt}
%the eigenvalues of the system matrix $\mathbb A_k$ are , which implies that the complex conjugate eigenvalues of $\mathbb A_k$ have negative real part for all $k\in\mathcal K$.

Under Assumption~\ref{assum4}, the eigenvalues of the system matrix  {$\mathbb A_k$} are computed as $\psi_k\pm\sqrt{\psi_k^2-\alpha_1^k\alpha_2^k(a_{11}^1a_{22}^2-a_{12}^1a_{12}^2)}
$, where $\psi_k\triangleq\frac{1}{2}(\alpha_1^ka_{11}^1+\alpha^k_2a_{22}^2)<0$, which implies that the complex conjugate eigenvalues of {$\mathbb A_k$} have negative real part for all $k\in\mathcal K$. Note that the expression in the square root satisfies
$
0>\psi_k^2-\alpha_1^k\alpha_2^k(a_{11}^1a_{22}^2-a_{12}^1a_{12}^2)=\frac{1}{4}(\alpha_1^ka_{11}^1-\alpha_2^ka_{22}^2)^2+\alpha_1^k\alpha_2^ka_{12}^1a_{12}^2,
$
which implies that $a_{12}^1a_{12}^2<0$ (i.e., $a_{12}^1<0\wedge a_{12}^2>0$ or $a_{12}^1>0\wedge a_{12}^2<0$) and
$\det  P_k =-\frac{1}{4}(\alpha_1^ka_{11}^1-\alpha_2^ka_{22}^2)^2-\alpha_1^k\alpha_2^ka_{12}^1a_{12}^2>0$, $k\in\mathcal K$.
These facts are used in the following lemma and its proof.
% {Under Assumption~\ref{assum4}, all the subsystems are stable \cite{yan2019Bifurcation} and $a_{12}^1a_{12}^2<0$ (i.e., $a_{12}^1<0\wedge a_{12}^2>0$ or $a_{12}^1>0\wedge a_{12}^2<0$~holds.}
 {Note that the case where $\mathbb A_k$ possesses real eigenvalues may also exhibit infinitely many mode transitions. This complicated case is addressed in \cite{yan2021loss}.}
%  {Recalling $a_{11}^1a_{22}^2>0$, we note that Assumption~\ref{assum4}
% implies $a_{12}^1a_{12}^2<0$ (i.e., $a_{12}^1<0\wedge a_{12}^2>0$ or $a_{12}^1>0\wedge a_{12}^2<0$).
%This fact is} used in the following statements.

\vspace{-5pt}
\begin{lemma}\label{cor1}
Consider the loss-aversion-based
noncooperative system $\mathcal G(J)$ with the pseudo-gradient dynamics (\ref{eqdynamics1}), (\ref{eqdynamics2})
%. If each $\mathbb A_k$ has a pair of complex conjugate eigenvalues for $k\in\mathcal K$,
under Assumption~\ref{assum4}. Then, the rotational directions of the trajectories are consistently the same in the entire state space $\mathbb R^2$. Specifically, if $a_{12}^1<0$ and $ a_{12}^2>0$ (resp., $a_{12}^1>0$ and $ a_{12}^2<0$), then
the trajectory
of (\ref{eqdynamics1}), (\ref{eqdynamics2}), keeps the counterclockwise (resp.,
clockwise) direction for any $\alpha_i^{\rm H}\geq\alpha_i^{\rm L}$, $i=1,2$.
\end{lemma}
\vspace{-6pt}

   {This result is used in the following sections to partition the state space and to define a piecewise linearized system of (\ref{eqdynamics1}), (\ref{eqdynamics2}). %among the overlapped regions (between $\mathcal D_k,k\in\mathcal K$) for defining the special system $\mathcal S$.
%As we show in the 3 cases below, the location of the Nash equilibrium matters since whether $A_i x^*+b_i$ vanishes or not exhibits hyperbolic/elliptic boundaries or simple straight lines for the partitioning.
}

\vspace{-9pt}
\subsection{Case 1: $A_ix^*+b_i\not=0$, $i\in\{1,2\}$}\label{section_B}
\vspace{-2pt}

In this section, we characterize the local stability property of the Nash equilibrium $x^*$ for $A_ix^*+b_i\not=0$ for $i\in\{1,2\}$.
 {Recall that $x^*$ is located on the cusp of the domains $\mathcal D_k$, $k \in \mathcal K$ (see Fig.~\ref{Region_Four})}.
In the beginning, we approximate the domain ${\mathcal D}_k$ around $x^*$ to the convex cone $\hat{\mathcal D}_k$ by linearizing the quadratic curves characterized by $\dot J^k_1(x)=0$ and $\dot J^k_2(x)=0$ around $x^*$ for all $k\in\mathcal K$. In particular, since $x^*$ corresponds to the origin in the shifted space $\tilde x$, we denote the linearized straight lines of the curves
$\dot J_i^k(x)=0$, $i\in\{1,2\}$, $k\in\mathcal K$, at $x^*$ as {\setlength\abovedisplayskip{1pt}
\setlength\belowdisplayskip{1pt}
\begin{eqnarray}\label{aaaa}
\gamma_i^k \tilde x_1+\tilde x_2=0,\quad i\in\{1,2\},\quad k\in\mathcal K,
\end{eqnarray}
where} $\gamma_i^k\triangleq
\big(\frac{\partial \dot J_i^k(x)}{\partial x_1}\big/
\frac{\partial \dot J_i^k(x)}{\partial x_2}\big)\big|_{x=x^*}\!\!\!\in\mathbb R$, $i\in\{1,2\}$, $k\in\mathcal K$. For example, Fig.~\ref{Region3} shows the domain $\mathcal D_{\rm LL}$ and its approximated cone $\hat{\mathcal D}_{\rm LL}$ in the neighborhood of $x^*$.
%\begin{figure}
%	\begin{center}
%		\includegraphics[width=8.5cm]{approximated_new-eps-converted-to.pdf}    % The printed column width is 8.4 cm.
%		\vspace{-5pt}\caption{Approximated domain where $a_{12}^2(a_{12}^1 x_1^*+a_{22}^1x_2^*+b_2^1)>0$, $a_{11}^1(a_{11}^2 x_1^*+a_{12}^2x_2^*+b_1^2)>0$. (a): ${\mathcal D}_{\rm LL}$, (b): the approximated domains $\hat{\mathcal D}_k$, $k\in\mathcal K$, around the neighborhood of Nash equilibrium. The rotational direction is counterclockwise since $\gamma_1^{\rm LL}<0\wedge\gamma_2^{\rm LL}>0$ implies $a_{12}^1<0\wedge a_{12}^2>0$.}
%\vspace{-18pt}
%		\label{Region3}
%	\end{center}
%\end{figure}

% {Before we present the stability results, we show some general characteristics of the approximated cones $\hat{\mathcal D}_k$, $k\in\mathcal K$.} %including the facts that all of the 4 approximated cones $\hat{\mathcal D}_k$, $k\in\mathcal K$, must exist; they do not possess any overlap with each other; the values of $\alpha_1^{\rm H},\alpha_1^{\rm L},\alpha_2^{\rm H},\alpha_2^{\rm L}$ do not affect the partition of $\hat{\mathcal D}_k$, $k\in\mathcal K$.
%To reveal such characteristics, the first primary concern is on the boundaries of $\hat{\mathcal D}_k$, $k\in\mathcal K$.
%The following result shows $\gamma_i^{\rm LL}=\gamma_i^{\rm HL}=\gamma_i^{\rm LH}=\gamma_i^{\rm HH}$, $i\in\{1,2\}$.
For the statement of the following result, note that $a_{12}^1\not=0$ and $a_{12}^2\not=0$ since $a_{12}^1a_{12}^2<0$ under Assumption~\ref{assum4}.
\vspace{-5pt}
\begin{prop}\label{slope}
If $A_ix^*+b_i\not=0$ for $i=1$ (resp., $i=2$), then
$\gamma_1^k=\frac{a_{12}^2}{a_{22}^2}$ (resp., $\gamma_2^k=\frac{a_{11}^1}{a_{12}^1}$), $k\in\mathcal K$, for any $\alpha_1^{\rm H},\alpha_1^{\rm L},\alpha_2^{\rm H},\alpha_2^{\rm L}\in\mathbb R_+$.
\end{prop}
\vspace{-5pt}

\vspace{-4pt}
\begin{rem}\emph{
 {It is  interesting to note from Proposition 1 that the linearized line (\ref{aaaa}) of $\dot J_1^k(x)=0$ coincides with the best-response line (\ref{approximated}) for agent 2 (instead of agent 1).
The similar observations hold for the linearized lines of $\dot J_2^k(x)=0$.}
}
\end{rem}
\vspace{-5pt}

\vspace{-5pt}
\begin{rem}\emph{Since $a_{11}^1a_{22}^2\not=a_{12}^1a_{12}^2$ holds in (\ref{eqqudractic}), it follows that $\gamma_1^k \not=\gamma_2^k$, $k\in\mathcal K$, and hence the boundaries of $\hat {\mathcal D}_{k}$, $k\in\mathcal K$, are simply characterized by the two \emph{intersected} straight lines
 {(\ref{approximated2}) and (\ref{approximated}).}
Consequently, since ${\rm int}\, \hat{\mathcal D}_k=\emptyset$ holds only for $\gamma_1^k=\gamma_2^k$, $k\in\mathcal K$, all of the 4 approximated cones must exist
 {with ${\rm int}\, {\mathcal D}_k$, $k\in\mathcal K$, being non-empty. }
}
\end{rem}
\vspace{-5pt}

\vspace{-6pt}
\begin{lemma}\label{lemma_4}
 {The approximated domains $\hat{\mathcal D}_k$, $k\in\mathcal K$, are identified to be the 4 convex cones partitioned by the best-response lines (\ref{approximated2}) and (\ref{approximated}), and satisfy $({\rm int }\, \hat{\mathcal D}_i)\cap ({\rm int }\,\hat{\mathcal D}_j)=\emptyset$ for $i,j\in\mathcal K$, $i\not=j$, ${\rm int}\, \hat{\mathcal D}_k\not=\emptyset$, $k\in\mathcal K$, for any $\alpha_i^{\rm H}\geq\alpha_i^{\rm L}$, $i=1,2$. Moreover, the domain $\hat{\mathcal D}_{\rm LL}$ (resp., $\hat{\mathcal D}_{\rm HL}$) is centrally symmetric about the Nash equilibrium $x^*$ to $\hat{\mathcal D}_{\rm HH}$ (resp., $\hat{\mathcal D}_{\rm LH}$). }
\end{lemma}\vspace{-3pt}

\vspace{-2pt}

\vspace{-6pt}
\begin{rem}\label{rem_phase}\emph{
 {Lemma~\ref{lemma_4} implies that the best-response lines~(\ref{approximated2}) and (\ref{approximated}) coincide with the switching phases (see a typical example of the approximated domains $\hat{\mathcal D}_k$, $k\in\mathcal K$, shown in Fig.~\ref{Region3}(b)) and hence} the switching phases at which agents switch the modes around the Nash equilibrium $x^*$ are given {by $\theta=\arctan(-\frac{a_{12}^2}{a_{22}^2} )$, $\arctan(-\frac{a_{11}^1}{a_{12}^1})$, $\arctan(-\frac{a_{12}^2}{a_{22}^2} )+\pi$,} $\arctan(-\frac{a_{11}^1}{a_{12}^1})+\pi$.
Recalling the fact shown in Lemma~\ref{cor1}, the transition of agents' modes around $x^*$ includes only two possibilities depending on the rotational directions, which are
%{\setlength\abovedisplayskip{1.5pt}
%\setlength\belowdisplayskip{1.5pt}\begin{eqnarray} \label{type1}
%&
{$\cdots\to{\rm HH}\to{\rm LH}\to{\rm LL}\to{\rm HL}\to{\rm HH}\to\cdots$ and %&\\
$\cdots\to{\rm HH}\to{\rm HL}\to{\rm LL}\to{\rm LH}\to{\rm HH}\to\cdots.$}
The transition sequence of the noncooperative system $\mathcal G(J)$ used for Fig.~\ref{Region3}(b) is depicted in Fig.~\ref{transition_ex}, where the sequence is given by {the former one} %(\ref{type1})
since the rotational direction of the trajectories is counterclockwise.
}\end{rem}
\vspace{-6pt}

\begin{figure*}
\begin{minipage}{0.32\textwidth}
\centering
		\includegraphics[width=5.8cm]{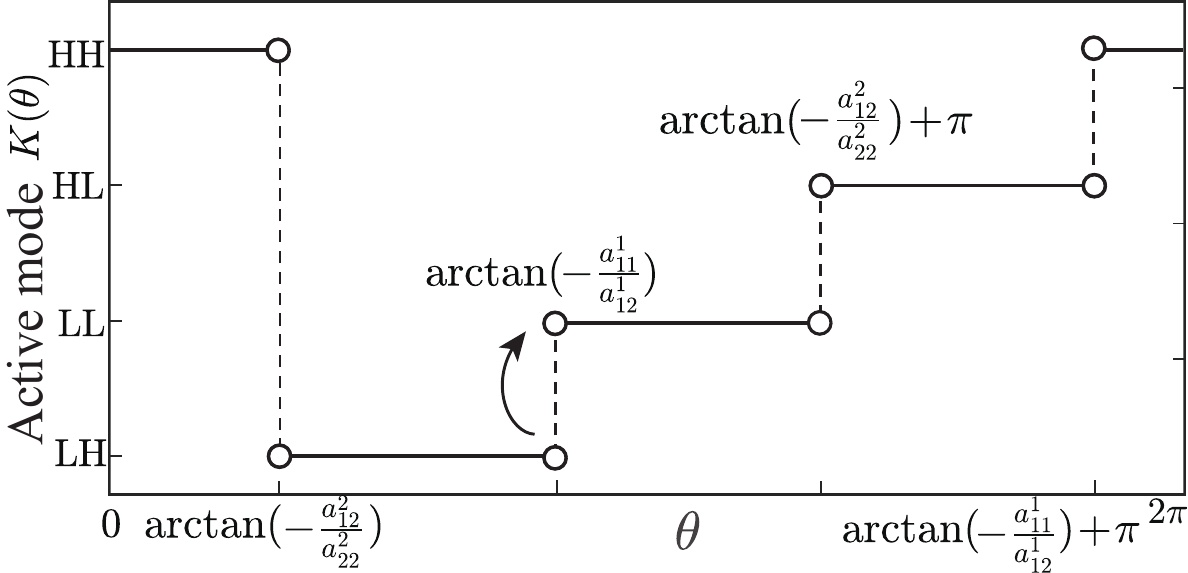}    % The printed column width is 8.4 cm.
\caption{Mode transition in (\ref{eqdynamics1}), (\ref{eqdynamics2}) around $x^*$ in the same $\mathcal G(J)$ as Fig.~\ref{Region3} where $A_ix^*+b_i\not=0$, $i\in\{1,2\}$.}
		\label{transition_ex}
\end{minipage}\hspace{2pt}
\begin{minipage}{0.354\textwidth}
\centering
		\includegraphics[width=5.4cm]{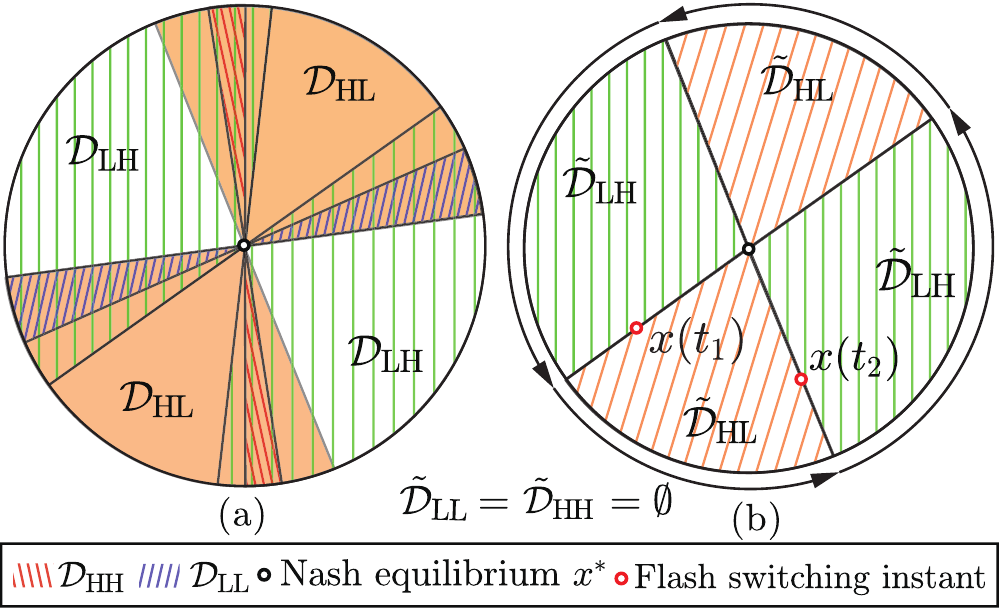}    % The printed column width is 8.4 cm.
		\caption{An example of the partition of $\tilde{\mathcal D}_k,k\in\mathcal K$, from the domains ${\mathcal D}_k,k\in\mathcal K$, for $A_ix^*+b_i=0,i\in\{1,2\}$. (a): ${\mathcal D}_k,k\in\mathcal K$, (b): effective domains $\tilde{\mathcal D}_k,k\in\mathcal K$, with counterclockwise trajectories.}
		\label{Region2}
\end{minipage}\hspace{2pt}
\begin{minipage}{0.31\textwidth}
\centering
		\includegraphics[width=5.5cm]{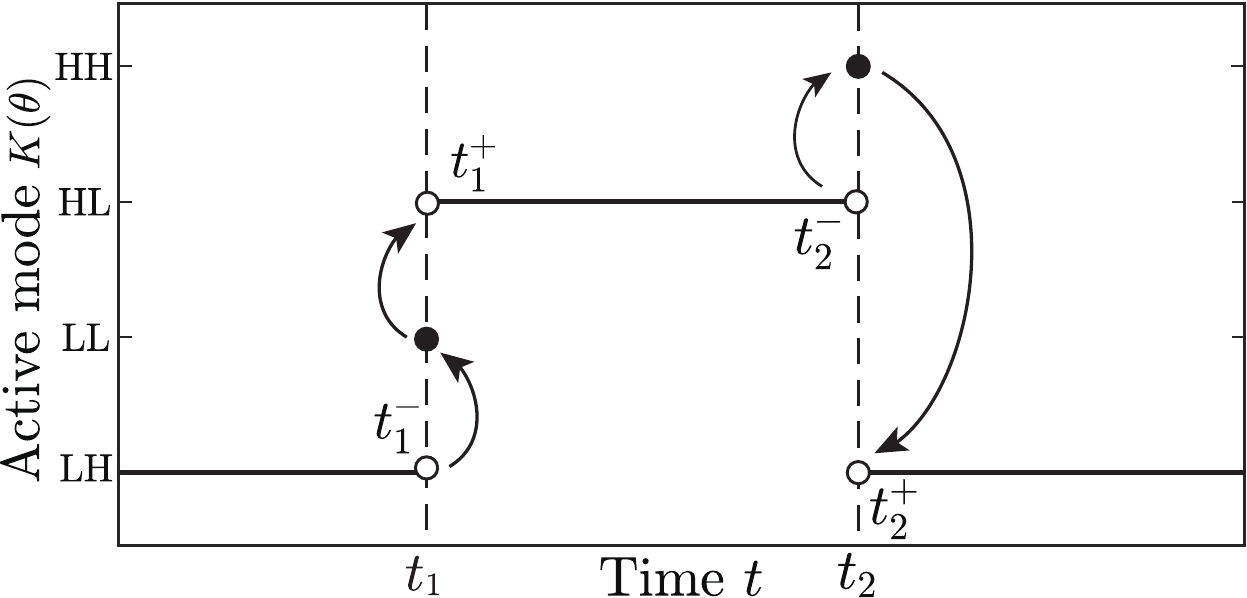}    % The printed column width is 8.4 cm.
		\caption{Mode transition sequence in (\ref{eqdynamics1}), (\ref{eqdynamics2}) around the flash switching instants $t_1$ and $t_2$ in Fig.~\ref{Region2}. At those flash switching instants, agent 2 first switches its sensitivity parameter and agent 1 further switches its sensitivity right
after agent 2's switch.}
		\label{transition}
\end{minipage}\vspace{-17pt}
\end{figure*}

%
%\begin{figure}
%	\begin{center}
%		\includegraphics[width=6cm]{transition_ex2-eps-converted-to.pdf}    % The printed column width is 8.4 cm.
%\vspace{-9pt}\caption{Mode transition in (\ref{eqdynamics1}), (\ref{eqdynamics2}) around $x^*$ in the same $\mathcal G(J)$ as Fig.~\ref{Region3} where $A_ix^*+b_i\not=0$, $i\in\{1,2\}$.}
%		\label{transition_ex}
%	\end{center}
%\vspace{-22pt}
%\end{figure}

Now, the local stability property of Nash equilibrium $x^*$ of the pseudo-gradient dynamics (\ref{eqdynamics1}), (\ref{eqdynamics2}) is equivalent to the stability property of the  {piecewise linearized system} given by
{\setlength\abovedisplayskip{2pt}
\setlength\belowdisplayskip{2pt}\begin{equation}\label{PLS}
 \dot{ x}(t)%&=\alpha^k\left[\frac{ \partial J_1( x(t))}{\partial  x_1},\frac{ \partial J_2( x(t))}{\partial  x_2}\right]^{\mathrm T}\\
 =\mathbb A_k( x(t)-x^*),\quad  x(t)\in\hat{\mathcal D}_k.
\end{equation}
Recalling} that $\hat{\mathcal D}_k, k\in\mathcal K$, satisfy $\bigcup_{k\in\mathcal K}\hat{\mathcal D}_k=\mathbb R^2$ (Lemma~\ref{lemma_4}) and $({\rm int }\, \hat{\mathcal D}_i)\cap ({\rm int }\,\hat{\mathcal D}_j)=\emptyset$ for $i,j\in\mathcal K$, $i\not=j$, we use the method shown in \cite{Nishiyama2008Optimal} to determine stability of the piecewise linear system (\ref{PLS}).
Specifically, define the normalized radial growth rate for each mode $k\in\mathcal K$ by\vspace{-1pt}
{\setlength\abovedisplayskip{1pt}
\setlength\belowdisplayskip{1pt}\begin{align}\label{rho_k}
\rho_k(\theta)
\triangleq\frac{1}{r}\frac{{\rm d}r}{{\rm d}\theta}=\frac{\eta^{\mathrm T}(\theta)\mathbb A_k\eta(\theta)}{\det[\eta(\theta),\mathbb A_k\eta(\theta)]}
=\frac{\eta^{\mathrm T}(\theta)\mathbb A_k\eta(\theta)}{\eta^{\mathrm T}(\theta)P_k\eta(\theta)},
\end{align}
where} $P_k$ is defined in (\ref{PK}). Note that $\rho_k(\theta)$, $k\in\mathcal K$, are continuous in $\theta$.
Then, the integral of the normalized radial growth rate is given by%\vspace{-3pt}
{\setlength\abovedisplayskip{1pt}
\setlength\belowdisplayskip{1pt}\begin{equation}\label{gamma}
 {\gamma_{\rm rg}}\triangleq\int_{\theta_0}^{\theta_0+2\pi}\rho_{K(\theta)}(\theta){\rm d} \theta,\vspace{-4pt}
\end{equation}
where}
%$\rho(\theta)=\rho_{K(\theta)}(\theta)$ and
$\theta_0\in\mathbb R$ and $K(\theta)\in\mathcal K$ is a function of the phase $\theta$ representing which
mode is active for (\ref{PLS}) around the Nash equilibrium $x^*$.
 {Note that $\gamma_{\rm rg}$ in (\ref{gamma}) is invariant under $\theta_0$ because $\rho_K(\theta)$ is a periodic function of $\theta$ of period $2\pi$. The value of $\gamma_{\rm rg}$ is numerically evaluated once the active mode $K(\theta)$ is determined.}
%It is important to note that the result of determining the rotational direction in Corollary~\ref{cor1} still hold in this section.
%That is to say, the rotational direction in  (\ref{eqdynamics1}), (\ref{eqdynamics2}) is counterclockwise (resp., clockwise) if $a_{12}^1<0\ \&\ a_{12}^2>0$ (resp., $a_{12}^1>0\ \&\ a_{12}^2<0$) and Assumptions~\ref{assum4} hold.

\vspace{-6pt}
\begin{thm}\label{thm2}
Consider the loss-aversion-based
noncooperative system $\mathcal G(J)$ with the pseudo-gradient dynamics (\ref{eqdynamics1}), (\ref{eqdynamics2}) under Assumption~\ref{assum4} for $A_ix^*+b_i\not=0$, $i\in\{1,2\}$.
If $a_{12}^1\gamma_{\rm rg}>0$ and $ a_{12}^2\gamma_{\rm rg}<0$ (resp., $a_{12}^1\gamma_{\rm rg}<0$ and $ a_{12}^2\gamma_{\rm rg}>0$), then the Nash equilibrium $x^*$ in (\ref{eqdynamics1}), (\ref{eqdynamics2}) hold, is asymptotically stable (resp., unstable).
\end{thm}
\vspace{-7pt}

\vspace{-4pt}
\begin{rem} \emph{
 {Even though it follows from Lemma~\ref{lemma_4} that} $\alpha_1^{\rm H},\alpha_1^{\rm L},\alpha_2^{\rm H}$, $\alpha_2^{\rm L}$ do not change the partition of $\hat{\mathcal D}_k$, $k\in\mathcal K$, they affect the normalized
radial growth rates $\rho_k$, $k\in\mathcal K$, in  {(\ref{rho_k})} by altering $P_k$, $\mathbb A_k$, $k\in\mathcal K$, and hence may change the stability property.}
\end{rem}

\vspace{-11pt}
\begin{rem} \emph{
The parameters
$a_{22}^1,a_{11}^2,b_2^1,b_1^2$ neither change the normalized
radial growth rates $\rho_k(\theta)$, $k\in\mathcal K$, nor the switching phases $\theta=\arctan(-\frac{a_{12}^2}{a_{22}^2})$, $\arctan(-\frac{a_{11}^1}{a_{12}^1} )$, $\arctan(-\frac{a_{12}^2}{a_{22}^2})+\pi$, $\arctan(-\frac{a_{11}^1}{a_{12}^1} )+\pi$,
but they affect the active mode $K(\theta)$ due to a permutation of the locations of $\hat{\mathcal D}_k$, $k\in\mathcal K$, among the 4 convex cones partitioned by (\ref{approximated2}) and (\ref{approximated}), and hence may change the stability property.}
\end{rem}

\vspace{-14pt}
\subsection{Case 2: $A_ix^*+b_i=0$, $i\in\{1,2\}$}\label{section_C}\vspace{-2pt}
In this section, we characterize the stability property of the Nash equilibrium $x^*$ for $A_ix^*+b_i=0$ for $i\in\{1,2\}$.
In such a case, recall that the domains $\mathcal D_k,k\in\mathcal K$, are convex cones with $x^*$ being the center since $\dot J^k_i( x)=(x-x^*)^\mathrm{T}Q_i^k(x-x^*)$, $i\in\{1,2\}$, $k\in\mathcal K$, in (\ref{J_ik}).

Note that if $Q_i^k>0$, $i\in\{1,2\}$, $k\in\mathcal K$, then it follows that ${\rm int\,}\mathcal D_{\rm LL},{\rm int\,}\mathcal D_{\rm HL},{\rm int\,}\mathcal D_{\rm LH}=\emptyset$, and $\mathcal D_{\rm HH}=\mathbb R^2$ (Remark~\ref{rem_nonexistence_may}) so that there is no mode transition.
Henceforth, in this section for \emph{Case 2}, suppose that the matrices $Q_i^k={Q_i^k}^{\mathrm T}$, $i\in\{1,2\}$, $k\in\mathcal K$, are all sign-indefinite.
Under this condition, each of the domains $\mathcal D_k$, $k\in\mathcal K$, satisfies $\mathcal D_k\not=\mathbb R^2$ and the boundaries of the existing convex cones $\mathcal D_k$ characterized by $\dot J^k_1(x)=0$
and/or $\dot J^k_2(x)=0$ are given by the 2 lines out of the 4 lines% represented by
{\setlength\abovedisplayskip{1pt}
\setlength\belowdisplayskip{0.1pt}\begin{align}\label{boundary_line1}
\tilde \gamma_1^{k+}\tilde x_1+\tilde x_2=0,\quad \tilde \gamma_1^{k-}\tilde x_1+\tilde x_2=0,\\ \label{boundary_line2}
\tilde \gamma_2^{k+}\tilde x_1+\tilde x_2=0,\quad \tilde \gamma_2^{k-}\tilde x_1+\tilde x_2=0,
\end{align}
where}
{\setlength\abovedisplayskip{1pt}
\setlength\belowdisplayskip{1pt}\begin{eqnarray}\label{slope_linear}
\tilde \gamma_i^{k\pm}\!\!=\!\frac{{Q_i^k}_{(1,2)}\pm\!\sqrt{{Q_i^k}_{(1,2)}{Q_i^k}_{(1,2)}-{Q_i^k}_{(1,1)}{Q_i^k}_{(2,2)}}}{{Q_i^k}_{(2,2)}}\in\mathbb R,
\end{eqnarray}
 and} $Q_i^k(a,b)$ denotes the $(a,b)$th entry of $Q_i^k$. %that is,
%{\setlength\abovedisplayskip{1pt}
%\setlength\belowdisplayskip{1pt}\begin{align}
%%    Q_1^k&=&\left[\begin{array}{cc}
%%               Q_1^k(1) & Q_1^k(2) \\
%%               Q_1^k(2) & Q_1^k(3)
%%             \end{array}\right],\\
%%    Q_2^k&=&\left[\begin{array}{cc}
%%               Q_2^k(1) & Q_2^k(2) \\
%%               Q_2^k(2) & Q_2^k(3)\end{array}\right],\\
%   & {Q_1^k}_{(1,1)}=2(\alpha^k_1a_{11}^1a_{11}^1+\alpha^k_2a_{12}^1a_{12}^2), \\
%   & {Q_1^k}_{(1,2)}=2\alpha^k_1a_{11}^1a_{12}^1+\alpha^k_2(a_{12}^1a_{22}^2+a_{12}^2a_{22}^1),\\
%   & {Q_1^k}_{(2,2)}=2(\alpha^k_1a_{12}^1a_{12}^1+\alpha^k_2a_{22}^1a_{22}^2),\\
%   & {Q_2^k}_{(1,1)}=2(\alpha^k_1a_{11}^1a_{11}^2+\alpha^k_2a_{12}^2a_{12}^2),\\
%   & {Q_2^k}_{(1,2)}=\alpha^k_1(a_{12}^1a_{11}^2+a_{12}^2a_{11}^1)+2\alpha^k_2a_{22}^2a_{12}^2,\\
%   & {Q_2^k}_{(2,2)}=2(\alpha^k_1a_{12}^1a_{12}^2+\alpha^k_2a_{22}^2a_{22}^2).
%\end{align}
%}
%\vspace{-6pt}

%\vspace{-6pt}
In general, it turns out that there may be overlapped regions between ${\mathcal D}_k$, $k\in\mathcal K$.
{Depending on the rotational direction characterized in Lemma~\ref{cor1},} we
define the effective domains~$\tilde {\mathcal D}_k$, $k\in\mathcal K$, indicating that which mode is active  in the overlapped regions by properly partitioning the state space. Specifically, we assume that the modes do not change until increasing/decresing property of $J_i$ changes so that agent $i$ switches its sensitivity parameter $\alpha_i(\cdot)$  {when agent $i$ reaches the boundary of the current mode} (see the effective domains  {for a trajectory moving in the counterclockwise direction} in Fig.~\ref{Region2}(b) yielded from the domains $\mathcal D_k$, $k\in\mathcal K,$ given in Fig.~\ref{Region2}(a)). Note as a direct consequence of Lemma~\ref{thm_nonzeno} that $\tilde{\mathcal D}_k$, $k\in\mathcal K$, satisfy $\bigcup_{k\in\mathcal K}\tilde{\mathcal D}_k=\mathbb R^2$.
%\begin{figure}
%	\begin{center}
%		\includegraphics[width=5.59cm]{Region_D_k5-eps-converted-to.pdf}    % The printed column width is 8.4 cm.
%		\vspace{-8pt}\caption{An example of the partition of $\tilde{\mathcal D}_k,k\in\mathcal K$, from the domains ${\mathcal D}_k,k\in\mathcal K$, under Assumption~\ref{assum4} for $A_ix^*+b_i=0,i\in\{1,2\}$. (a): ${\mathcal D}_k,k\in\mathcal K$, (b): effective domains $\tilde{\mathcal D}_k,k\in\mathcal K$, with counterclockwise trajectories.
%\vspace{-28pt}
%}
%		\label{Region2}
%	\end{center}
%\end{figure}
%

Consequently, the stability property of the Nash equilibrium $x^*$ of the pseudo-gradient dynamics (\ref{eqdynamics1}), (\ref{eqdynamics2}) is equivalent to the stability property in the piecewise linear system given by (\ref{PLS}) with $\hat{\mathcal D}_k$ replaced by $\tilde{\mathcal D}_k$.
%where $\tilde{\mathcal D}_k, k\in\mathcal K$, satisfy $\bigcup_{k\in\mathcal K}\tilde{\mathcal D}_k=\mathbb R^2$ and $({\rm int}\, \tilde{\mathcal D}_i)\cap ({\rm int}\,\tilde{\mathcal D}_j)=\emptyset$ for $i,j\in\mathcal K$, $i\not=j$.
Similar to the previous section, we use the integral of the normalized radial growth rate $\gamma_{\rm rg}$ to determine stability of the Nash equilibrium $x^*$. Note that since the active mode of (\ref{PLS}) at phase $\theta+\pi$ is exactly same as the active mode at phase $\theta$ (i.e., $K(\theta+\pi)=K(\theta)$), we have
$\gamma_{\rm rg}=2\int_{\theta_0}^{\theta_0+\pi}\rho_{K(\theta)}(\theta){\rm d} \theta$.

\vspace{-5pt}
\begin{thm}\label{thm1}
Consider the loss-aversion-based
noncooperative system $\mathcal G(J)$ with the pseudo-gradient dynamics (\ref{eqdynamics1}), (\ref{eqdynamics2}) under Assumption~\ref{assum4} for $A_ix^*+b_i=0,i\in\{1,2\}$.
Then the following statements hold:

1) If $a_{12}^1\gamma_{\rm rg}>0$ and $ a_{12}^2\gamma_{\rm rg}<0$, then the Nash equilibrium $x^*$ in (\ref{eqdynamics1}), (\ref{eqdynamics2}) is globally asymptotically stable;

2) If $\gamma_{\rm rg}=0$, then (\ref{eqdynamics1}), (\ref{eqdynamics2}) are marginally stable and the trajectory
of (\ref{eqdynamics1}), (\ref{eqdynamics2}) constitutes a closed orbit;

3) If $a_{12}^1\gamma_{\rm rg}<0$ and $ a_{12}^2\gamma_{\rm rg}>0$, then the Nash equilibrium $x^*$ in (\ref{eqdynamics1}), (\ref{eqdynamics2}) is unstable.
\end{thm}
%\vspace{-10pt}
%
%
%\begin{rem} \emph{
% {Since
%$\tilde \gamma_i^{k\pm}$ defined in (\ref{slope_linear}) is probably neither $\frac{a_{12}^2}{a_{22}^2}$ nor $\frac{a_{11}^1}{a_{12}^1}$, the switching phases may not be $\arctan(-\frac{a_{12}^2}{a_{22}^2} )$, $\arctan(-\frac{a_{11}^1}{a_{12}^1})$, $\arctan(-\frac{a_{12}^2}{a_{22}^2} )+\pi$, $\arctan(-\frac{a_{11}^1}{a_{12}^1})+\pi$}.}
%\end{rem}

\vspace{-5pt}

Next, we present several interesting observations on agents' behavior in the following statements.
In terms of the mode transition sequence, there may exist some time instant $t$ at phase $\theta$ at which the agents switch the sensitivity parameters such that the active mode $K(\theta(t))$ experiences
{\setlength\abovedisplayskip{1pt}
\setlength\belowdisplayskip{1pt}\begin{align}
K(\theta(t^-))\not=K(\theta(t))\not=K(\theta(t^+)).
\end{align}
We} call such a switching instant $t$ as a \emph{flash switching instant}.
Figure~\ref{transition} shows an example of the mode transition around a \emph{flash switching instant} $t_1$ used for Fig.~\ref{Region2}(b), where agents' state enters into $\mathcal D_{\rm HL}$ after leaving $\mathcal D_{\rm LH}$ at time $t_1$. In this example,  {when the 2 agents are in the domain $\mathcal D_{\rm LH}$ and agent 2 reaches its boundary at $t_1$}, agent 2 switches the sensitivity from $\alpha_2^{\rm H}$ to $\alpha_2^{\rm L}$ since  {$\dot J_2(t_1)$ becomes 0 from $\dot J_2(t_1^-)>0$}.
However, since agent 2's switching behavior results in  {$\dot J_1(t_1)>0$ from $\dot J_1(t_1^-)<0$},
agent 1 further switches its sensitivity from $\alpha_1^{\rm L}$ to $\alpha_1^{\rm H}$ right after the agent 2's switch  ($t_1^+$).
After time $t_1^+$, since agents' state successfully enters into the domain $\mathcal D_{\rm HL}$, the agents keep the mode $\rm HL$.
In the example of Fig.~\ref{Region2}, the next switching instant $t_2$ (and all the switching instants) shown to be flash   switching instants as well. In short, the reason why there may exist a {flash switching instant} is that
a single agents' sensitivity transition can  {be a trigger to make} the other agent immediately switch its sensitivity almost at the same time instant.
%The moment of such phenomenon can be understood as a \emph{flash switching point} in agents' mode transition sequence.
%

%\begin{figure}
%	\begin{center}
%		\includegraphics[width=5.5cm]{no_solution-eps-converted-to.pdf}    % The printed column width is 8.4 cm.
%	\vspace{-9pt}	\caption{Mode transition sequence in (\ref{eqdynamics1}), (\ref{eqdynamics2}) around the flash switching instants $t_1$ and $t_2$ in Fig.~\ref{Region2}. At those flash switching instants, agent 2 first switch its sensitivity parameter and agent 1 further switches its sensitivity right
%after agent 2's switch.
%\vspace{-25pt}
%%As a result, at $t_1$, agents leave $\mathcal D_{\rm LH}$ and enter into $\mathcal D_{\rm HL}$; at $t_2$, agents leave $\mathcal D_{\rm HL}$ and enter into $\mathcal D_{\rm LH}$.
%}
%		\label{transition}
%	\end{center}
%\end{figure}

\begin{figure*}
\begin{minipage}{0.4\textwidth}
\centering
\includegraphics[width=7.5cm]{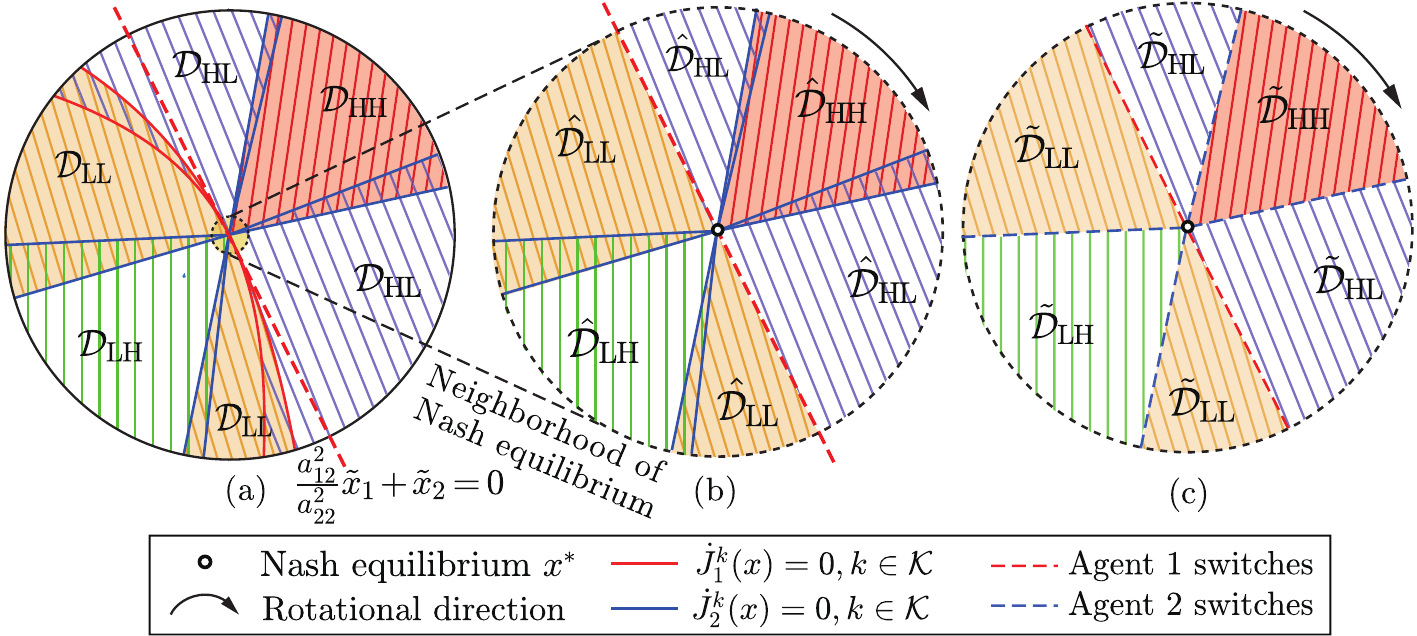}    % The printed column width is 8.4 cm.
	 \caption{An example of the partitions of ${\mathcal D}_k,k\in\mathcal K,\hat{\mathcal D}_k,k\in\mathcal K,$ and $\tilde{\mathcal D}_k,k\in\mathcal K$, under Assumption~\ref{assum4} for $A_1x^*+b_1\not=0,A_2x^*+b_2=0$. (a): ${\mathcal D}_k,k\in\mathcal K$, (b): approximated domains $\hat{\mathcal D}_k,k\in\mathcal K$, (c): effective domains $\tilde{\mathcal D}_k,k\in\mathcal K$, determined from (b) with trajectories moving in the clockwise direction.
}\label{Region4}
\end{minipage} \hspace{3pt}
\begin{minipage}{0.584\textwidth}
\centering
\includegraphics[width=10.5cm]{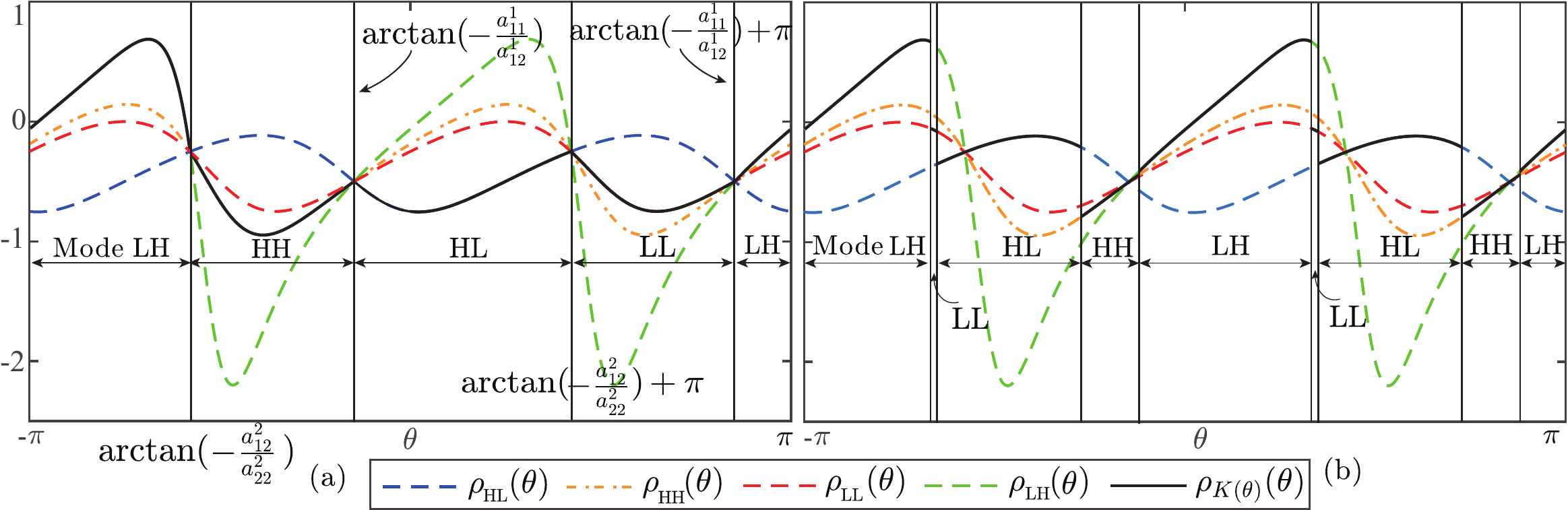}    % The printed column width is 8.4 cm.
		 \caption{Typical normalized
radial growth rates $\rho_{K(\theta)}(\theta),\rho_k(\theta),k\in\mathcal K,\theta\in[-\pi,\pi]$, with {the} same $A_1,A_2,b_1^1,b_2^2$ but different $b_1^2,b_2^1$. (a): $A_ix^*+b_i\not=0,i\in\{1,2\}$ (\emph{Case 1}), (b): $A_ix^*+b_i=0,i\in\{1,2\}$ (\emph{Case 2}). The parameters $b_1^2,b_2^1$ in (a) are obtained by giving small perturbations on $b_1^2$ and $b_2^1$ in (b).
}
		\label{rho}
\end{minipage} \vspace{-15pt}
\end{figure*}

Under the following assumption, the next results show that the effective domain $\tilde{\mathcal D}_{\rm LL}$ can never be adjacent to $\tilde{\mathcal D}_{\rm HH}$ and a \emph{flash switching instant} $t$ exists \emph{only if} the sequence of the active modes satisfies
%{\setlength\abovedisplayskip{1pt}\begin{eqnarray}\label{nece_flash}
{$(K(\theta(t^-)),K(\theta(t^+)))\in\{(\rm LH,HL),(HL,LH)\}$.}
%\end{eqnarray}}

\vspace{-6pt}
\begin{assum} \label{assum5} \emph{The straight lines characterized by $\dot J^{k}_1(x)=0$ do not coincide with the lines characterized by $\dot J^{k}_2(x)=0$ for any modes $k\in\mathcal K$.
In other words, $\tilde \gamma_1^{k+},\tilde \gamma_1^{k-},\tilde \gamma_2^{k+},\tilde \gamma_2^{k-}$ are all different in (\ref{boundary_line1}), (\ref{boundary_line2}) when $k$ is fixed.}\end{assum}
\vspace{-7pt}

%With a slight abuse of notation for $K(t)$ representing $K(\theta(t))$.
%
%Before we present a theorem, we give the following lemma.
%
%\vspace{-6pt}
%\begin{lemma}\label{lemma1}
%If both $A_1$ and $A_2$ in (\ref{eqqudractic}) are sign-indefinite under Assumption~\ref{assum4} for $A_ix^*+b_i=0,i\in\{1,2\}$, then ${\rm int}\, \mathcal D_{\rm HL}$ and ${\rm int}\,\mathcal D_{\rm LH}$ are non-empty for any $\alpha_1^{\rm H},\alpha_1^{\rm L},\alpha_2^{\rm H},\alpha_2^{\rm L}\in\mathbb R_+$.
%Furthermore, the best-response line $a_{11}^1  x_1+a_{12}^1 x_2+b_1^1=0$ for agent 1 (resp., $a_{12}^2  x_1+a_{22}^2 x_2+b_2^2=0$ for agent 2) belongs only to ${\rm int}\, \mathcal D_{\rm LH}$ (resp., ${\rm int}\, \mathcal D_{\rm HL}$).
%\end{lemma}
%\vspace{-5pt}

%\begin{rem}\emph{The typical example of a noncooperative system satisfying the condition in Lemma~\ref{lemma1} is a zero-sum game.
%}
%\end{rem}
\vspace{-5pt}
\begin{thm}\label{transition_thm}
Let $t_1,t_2$ be two consecutive switching instants for the noncooperative system $\mathcal G(J)$ under Assumption~\ref{assum5}  for $A_ix^*+b_i=0,i\in\{1,2\}$.
If $K(\theta(t))=\rm LL$ or $\rm HH$ for $t_1<t<t_2$, then neither the switching instant $t_1$ nor $t_2$ is a flash switching instant and the mode transition satisfies
$K(\theta(t_1^-)),K(\theta(t_2^+))\in\{\rm LH,HL\}$
for any $\alpha_i^{\rm H}\geq\alpha_i^{\rm L}$, $i=1,2$.
If, in addition, both $A_1$ and $A_2$ are sign-indefinite with Assumption~\ref{assum4}, then
 $(K(\theta(t_1^-)),K(\theta(t_2^+)))\in\{(\rm LH,HL),(HL,LH)\}$
for any $\alpha_i^{\rm H}\geq\alpha_i^{\rm L}$, $i=1,2$.
\end{thm}
\vspace{-6pt}

\vspace{-4pt}
\begin{rem}\emph{
Theorem~\ref{transition_thm} implies that if $\tilde{\mathcal D}_{\rm LL}$ or $\tilde{\mathcal D}_{\rm HH}$ exists for sign-indefinite $A_1$, $A_2$, then $\tilde{\mathcal D}_{\rm LH}$ and $\tilde{\mathcal D}_{\rm HL}$ are adjacent to $\tilde{\mathcal D}_{\rm LL}$ and/or $\tilde{\mathcal D}_{\rm HH}$ and hence the mode transition sequence around $\tilde{\mathcal D}_{\rm LL}$ and $\tilde{\mathcal D}_{\rm HH}$ is respectively given by
{$\cdots \rightleftharpoons\rm LH\rightleftharpoons LL\rightleftharpoons HL \rightleftharpoons\cdots$ and $
\cdots \rightleftharpoons\rm LH\rightleftharpoons HH\rightleftharpoons HL \rightleftharpoons\cdots$. }
%\cdots \rightleftharpoons\rm LH\rightleftharpoons LL\rightleftharpoons HL \rightleftharpoons\cdots$ and $
%\cdots \rightleftharpoons\rm LH\rightleftharpoons HH\rightleftharpoons HL \rightleftharpoons\cdots$.
Alternatively,
in the case where $\tilde{\mathcal D}_{\rm LL}=\tilde{\mathcal D}_{\rm HH}=\emptyset$ (as in Fig.~\ref{Region2}), both $\tilde{\mathcal D}_{\rm LH}$ and $\tilde{\mathcal D}_{\rm HL}$ must exist, since $\tilde{\mathcal D}_{\rm LH}\cup\tilde{\mathcal D}_{\rm HL}=\mathbb R^2$ and $\tilde{\mathcal D}_{ {\rm LH}},\tilde{\mathcal D}_{ {\rm HL}}\not=\mathbb R^2$. In such a case,
the mode transition sequence is given by {$
\cdots \rightleftharpoons\rm LH\rightleftharpoons HL \rightleftharpoons\cdots.$}
As a result, the modes $\rm LH$ and $\rm HL$ always exist when $A_1$ and $A_2$ are sign-indefinite.}
\end{rem}
%However, it is important to note that for the case where $A_1$ and $A_2$ are both negative indefinite or only one of them is sign-definite,
%$\rm LH$ or/and $\rm HL$  $Q_i^k,i\in\{1,2\},k\in\mathcal K$, are all sign-indefinite

\vspace{-10pt}
\begin{rem}\emph{
Theorem~\ref{transition_thm} does not imply that there always exists a flash switching instant when the mode transition $\rm LH\to HL$ or $\rm HL\rm\to LH$ happens.
For instance, consider the case with zero-sum payoffs. In this case, the overall mode transition sequence is composed of only modes $\rm LH$ and $\rm HL$ and the agents always simultaneously switch the sensitivity parameters at the same switching instants since the straight lines $\dot J_1^{\rm LH}(x)=0$ and $\dot J_2^{\rm LH}(x)=0$ (or, $\dot J_1^{\rm HL}(x)=0$ and $\dot J_2^{\rm HL}(x)=0$) coincide with each other. As a result, the switching instants in such a system are not flash switching instants.
%In fact, the switching instant $t$ is a flash switching instant if and only if $(K(\theta(t^-)),K(\theta(t^+)))\in\{(\rm LH,HL),(HL,LH)\}$ and $\dot J^{K(\theta(t^-))}_1(x(t))=0\wedge\dot J^{K(\theta(t^-))}_2(x(t))\not=0$ (or $\dot J^{K(\theta(t^-))}_1(x(t))\not=0\wedge\dot J^{K(\theta(t^-))}_2(x(t))=0$) hold.
}
\end{rem}
\vspace{-6pt}

Note that the case where $Q_1^k$, $k\in\mathcal K$, are positive definite and $Q_2^k$, $k\in\mathcal K$, are sign-indefinite can be similarly handled by evaluating the sign of $\gamma_{\rm rg}$ in (\ref{gamma}) with possibly fewer number of domains.
%Finally, when $Q_i^k,i\in\{1,2\},k\in\mathcal K$, are all positive definite, it follows that ${\mathcal D}_{ {\rm HH}}=\mathbb R^2$, ${\mathcal D}_{k}=\emptyset$, $k\in\{\rm LL,LH,HL\}$ (and hence $\tilde{\mathcal D}_{ {\rm HH}}=\mathbb R^2,\tilde{\mathcal D}_{ k}=\emptyset,k\in\{\rm LL,LH,HL\}$), which implies that there is no switching instant.

\vspace{-14pt}
\subsection{Case 3: $A_1x^*+b_1\not=0$, $A_2x^*+b_2=0$}\label{section_D}\vspace{-3pt}
In this section, we characterize the stability property of the Nash equilibrium $x^*$ for $A_1x^*+b_1\not=0$, $A_2x^*+b_2=0$ with all sign-indefinite matrices $Q_2^k$, $k\in\mathcal K$, in (\ref{J_ik}).
In such a case, each of the domains $\mathcal D_k$, $k\in\mathcal K$, is understood as the overlap of convex cones and the regions whose boundaries are characterized by hyperbolic (or elliptic) functions (see the example shown in Fig.~\ref{Region_LL}(c) for $k=\rm LL$).

 {Similar to \emph{Case 1} in Section~\ref{section_B}}, we approximate the domain ${\mathcal D}_k$ around $x^*$ to the convex cone $\hat{\mathcal D}_k$ by linearizing the quadratic curve characterized by $\dot J^k_1(x)=0$ around $x^*$  %As we analyzed in Section~\ref{section_B}, all of the 4 quadratic curves $\dot J^k_1(x)=0$, $k\in\mathcal K$, around $x^*$ for $A_1x^*+b_1\not=0$ are linearized by
to the straight line (\ref{approximated}) for all $k\in\mathcal K$.
%To show $\bigcup_{k\in\mathcal K}\hat{\mathcal D}_k=\mathbb R^2$,
%we recall $\Delta_2^2(x)\geq 0$ and $\delta_2=\alpha_2^{\rm H}-\alpha_2^{\rm L}\geq 0$.
%Then, it follows from (\ref{J_dot_sum}) that
%$\{ x\in\mathbb R^2:\dot J^{\rm LL}_2( x)> 0\}\subseteq\{ x\in\mathbb R^2:\dot J_2^{\rm LH}(x)=\dot J_2^{\rm LL}(x)+\delta_2\Delta_2^2(x)>0\}$ and $\{ x\in\mathbb R^2:\dot J^{\rm HH}_2( x)< 0\}\subseteq\{ x\in\mathbb R^2:\dot J_2^{\rm HL}(x)=\dot J_2^{\rm HH}(x)-\delta_2\Delta_2^2(x)<0\}$, and hence
%{\setlength\abovedisplayskip{2pt}
%\setlength\belowdisplayskip{2pt}\begin{eqnarray}\nonumber
%\mathbb R^2\!\!\!\!&=&\!\!\!\!\{ x\in\mathbb R^2\!:\!\dot J^{\rm LL}_2( x)> 0\}\cup  \{ x\in\mathbb R^2\!:\!\dot J_2^{\rm LL}(x)\leq 0\}\\ \label{tool1}
%&=&\!\!\!\!\{ x\in\mathbb R^2\!:\!\dot J^{\rm LH}_2( x)\geq 0\}\cup  \{ x\in\mathbb R^2\!:\!\dot J_2^{\rm LL}(x)\leq 0\},\quad\,\,\\ \nonumber
%\mathbb R^2\!\!\!\!&=&\!\!\!\!\{ x\in\mathbb R^2\!:\!\dot J^{\rm HH}_2( x)< 0\}\cup  \{ x\in\mathbb R^2\!:\!\dot J_2^{\rm HH}(x)\geq 0\}\\ \label{tool2}
%&=&\!\!\!\!\{ x\in\mathbb R^2\!:\!\dot J^{\rm HL}_2( x)\leq 0\}\cup  \{ x\in\mathbb R^2\!:\!\dot J_2^{\rm HH}(x)\geq 0\}.\quad\,\,
%\end{eqnarray}
%Recalling} that $\{ x\in\mathbb R^2\!:\!\dot J^{k}_1( x)\geq 0\}$, $k\in\mathcal K$, (or, $\{ x\in\mathbb R^2\!:\!\dot J^{k}_1( x)\leq 0\}$, $k\in\mathcal K$) around $x^*$ share exactly the same half plane,
%it follows from (\ref{tool1}) and (\ref{tool2}) that
 {It can be similarly shown that} $\hat{\mathcal D}_{\rm LL}\cup\hat{\mathcal D}_{\rm LH}$ and $\hat{\mathcal D}_{\rm HH}\cup\hat{\mathcal D}_{\rm HL}$ are the
two half planes partitioned by (\ref{approximated}) (see Fig.~\ref{Region4}(b)).
%Hence, $\hat{\mathcal D}_k$, $k\in\mathcal K$, satisfy $\bigcup_{k\in\mathcal K}\hat{\mathcal D}_k=\mathbb R^2$.
%The approximated domains of the domain ${\mathcal D}_k$ shown in Fig.~\ref{Region4} (a) are illustrated in Figure~\ref{Region4} (b).
%It can be seen from Fig.~\ref{Region4} (b) that there may exist some overlaps between the approximated domains $\hat{\mathcal D}_k$, $k\in\mathcal K$.
Then, considering the overlapped regions, similar to \emph{Case 2} (Section~\ref{section_C}),
we define the effective domains $\tilde {\mathcal D}_k$, $k\in\mathcal K$, by partitioning the state space according to the rotational direction (see Fig.~\ref{Region4}(c)).  {{Different from \emph{Case 2} where some of the effective domains $\tilde{\mathcal D}_{k},k\in\mathcal K$, may be empty,
none of $\tilde{\mathcal D}_{k}$, $k\in\mathcal K$, is empty in \emph{Case~3} and hence all of the 4 modes exist.
}}
%since $\tilde{\mathcal D}_{\rm LL}\cup\tilde{\mathcal D}_{\rm LH}=\hat{\mathcal D}_{\rm LL}\cup\hat{\mathcal D}_{\rm LH}$ and $\tilde{\mathcal D}_{\rm HH}\cup\tilde{\mathcal D}_{\rm HL}=\hat{\mathcal D}_{\rm HH}\cup\hat{\mathcal D}_{\rm HL}$ are the two half planes partitioned by (\ref{approximated}) and $\tilde{\mathcal D}_{k},k\in\mathcal K$, can never be a half plane when $Q_2^k$, $k\in\mathcal K$, are all sign-indefinite.
Then, %by noting $\bigcup_{k\in\mathcal K}\tilde{\mathcal D}_k=\mathbb R^2$ and $({\rm int}\, \tilde{\mathcal D}_i)\cap ({\rm int}\,\tilde{\mathcal D}_j)=\emptyset$ for $i,j\in\mathcal K$, $i\not=j$,
the stability property of the Nash equilibrium $x^*$ of the pseudo-gradient dynamics (\ref{eqdynamics1}), (\ref{eqdynamics2}) is equivalent to the stability property in the piecewise linearized system given by (\ref{PLS}) {with $\hat{\mathcal D}_k$ replaced by $\tilde{\mathcal D}_k$}.
%Moreover, %even for the case where $A_1,A_2$ are sign-indefinite,
%{unlike \emph{Case 2},} the transition sequence (\ref{sequnce1}), (\ref{sequnce2}) may not be true, since both the next and the previous modes of $\rm LL$ or $\rm HH$ can be the same (see the typical example where the mode transition sequence around $\tilde{\mathcal D}_{\rm HH}$ is given by $\rm HL\to HH \to\rm HL$ in Fig.~\ref{Region4}(c)). }}

\vspace{-6pt}
\begin{thm}\label{thm3}
Consider the loss-aversion-based
noncooperative system $\mathcal G(J)$ with the pseudo-gradient dynamics (\ref{eqdynamics1}), (\ref{eqdynamics2}) {under} Assumption~\ref{assum4} for $A_1x^*+b_1\not=0$, $A_2x^*+b_2=0$.
If $a_{12}^1\gamma_{\rm rg}>0$ {and} $ a_{12}^2\gamma_{\rm rg}<0$ (resp., $a_{12}^1\gamma_{\rm rg}<0$ {and} $ a_{12}^2\gamma_{\rm rg}>0$), then the Nash equilibrium $x^*$ in (\ref{eqdynamics1}), (\ref{eqdynamics2}) is asymptotically stable (resp., unstable), where $\gamma_{\rm rg}$ is defined in (\ref{gamma}).
\end{thm}
\vspace{-6pt}

%\begin{figure}
%	\begin{center}
%		\includegraphics[width=8.5cm]{Region_D_case3_new-eps-converted-to.pdf}    % The printed column width is 8.4 cm.
%		\vspace{-10pt}\caption{An example of the partitions of ${\mathcal D}_k,k\in\mathcal K,\hat{\mathcal D}_k,k\in\mathcal K,$ and $\tilde{\mathcal D}_k,k\in\mathcal K$, under Assumption~\ref{assum4} for $A_1x^*+b_1\not=0,A_2x^*+b_2=0$. (a): ${\mathcal D}_k,k\in\mathcal K$, (b): approximated domains $\hat{\mathcal D}_k,k\in\mathcal K$, (c): effective domains $\tilde{\mathcal D}_k,k\in\mathcal K$, from (b) with trajectories moving in clockwise direction.
%\vspace{-21pt}
%}
%		\label{Region4}
%	\end{center}
%\end{figure}

%To reveal the fact that there does not exist any \emph{flash switching instant} in the loss-aversion-based noncooperative system around the Nash equilibrium $x^*$ for $A_1x^*+b_1\not=0$, $A_2x^*+b_2=0$,
%we show the following result which is similar to Theorem~\ref{transition_thm} since it implies that $\tilde{\mathcal D}_{\rm LL}$ can never be adjacent to $\tilde{\mathcal D}_{\rm HH}$.
\vspace{-5pt}
\begin{prop}\label{prop2}
%Let $t_1,t_2$ be two consecutive switching instants.
%If $K(\theta(t))=\rm LL$ or $\rm HH$ for $t_1<t<t_2$, then the switching instant $t_1$ nor $t_2$ is a flash switching instant and $K(\theta(t_1^-))$, $K(\theta(t_2^+))\in\{\rm LH, HL\}$
%for any $\alpha_i^{\rm H}\geq\alpha_i^{\rm L}$, $i=1,2$.
{Assume that $x^*$ satisfies $A_1x^*+b_1\not=0$, $A_2x^*+b_2=0$. Then there is no flash switching instant
%If $K(\theta(t))=\rm LL$ or $\rm HH$ (resp., $\rm LH$ or $\rm HL$) for $t_1<t<t_2$, then  and $K(\theta(t_1^-))$, $K(\theta(t_2^+))\in\{\rm LH, HL\}$ (resp., $\{\rm LL, HH\}$)
for any $\alpha_i^{\rm H}\geq\alpha_i^{\rm L}$, $i=1,2$, in the neighborhood of $x^*$.}
\end{prop}
\vspace{-3pt}

%Then, note that there would exist a \emph{flash switching instant} around the Nash equilibrium $x^*$ only when $\tilde{\mathcal D}_{\rm HL}$ could be adjacent to $\tilde{\mathcal D}_{\rm LH}$. In such a case, agent 2 should immediately switch its sensitivity right after agent 1's switch. However,

%

\vspace{-2pt}
Note that the case where {$Q_2^k$}, $k\in\mathcal K$, are positive definite can be similarly handled by evaluating the sign of $\gamma_{\rm rg}$ in (\ref{gamma}) with possibly fewer number of domains. {In this case, }the modes $\rm LH$ and $\rm HH$ always exist.
As a result, there exist at least 2 modes in the loss-aversion-based noncooperative system for $A_1x^*+b_1\not=0$, $A_2x^*+b_2=0$.

 \vspace{-12pt}
\subsection{Discussion on Small Perturbation on the Parameters and Non-Quadratic Payoff Functions}\vspace{-2pt}
In this section, we further compare the {\emph{Cases 1--3}} characterized in the previous sections in terms {of the} normalized radial growth rate {and extend the proposed framework for the case where the order of the payoff functions is greater than 2.}

The following result shows a special property of the normalized
radial growth rates $\rho_k(\theta)$, $k\in\mathcal K$, defined in (\ref{rho_k}).
\vspace{-5pt}

\begin{prop}\label{no_jump}
The normalized
radial growth rates $\rho_k(\theta)$, $k\in\mathcal K$, possess the common values at the 4 phases $\textstyle
\theta=\arctan(-\frac{a_{12}^2}{a_{22}^2} )$,  $\textstyle \arctan(-\frac{a_{11}^1}{a_{12}^1} )$, $\arctan(-\frac{a_{12}^2}{a_{22}^2} )+\pi$, $\arctan(-\frac{a_{11}^1}{a_{12}^1} )+\pi$
{characterized as the switching phases for Case~1 (Remark~\ref{rem_phase})}.
Specifically,
{$
\rho_k(\arctan (-\frac{a_{12}^2}{a_{22}^2}))=\rho_{k}(\arctan (-\frac{a_{12}^2}{a_{22}^2})+\pi)=\frac{a_{22}^2}{a_{12}^2}$,
$\rho_k(\arctan (-\frac{a_{11}^1}{a_{12}^1}))=\rho_{k}(\arctan (-\frac{a_{11}^1}{a_{12}^1})+\pi)=-\frac{a_{11}^1}{a_{12}^1}
$}
for  all $k\in\mathcal K$ with any $\alpha_1^{\rm H},\alpha_1^{\rm L},\alpha_2^{\rm H},\alpha_2^{\rm L}\in\mathbb R_+$.
\end{prop}
\vspace{-6pt}

\vspace{-4pt}
\begin{rem}
\emph{
Proposition~\ref{no_jump} implies that the normalized
radial growth rate $\rho_{K(\theta)}(\theta)$ in \emph{Case~1} is continuous on $\theta$, since $\rho_k(\theta)$, $k\in\mathcal K$, possess the same {values} at the 4 switching phases
(see Fig.~\ref{rho}(a)).
However, in \emph{Cases~2} and~\emph{3}, since
agents may switch the sensitivity parameters at a phase $\theta\not\in\{\arctan(-\frac{a_{12}^2}{a_{22}^2} )$, $\arctan(-\frac{a_{11}^1}{a_{12}^1})$, $\arctan(-\frac{a_{12}^2}{a_{22}^2} )+\pi$, $\arctan(-\frac{a_{11}^1}{a_{12}^1})+\pi\}$,
$\rho_{K(\theta)}(\theta)$ is most {likely to be} discontinuous at the switching phases (see Fig.~\ref{rho}(b)).
}
\end{rem}\vspace{-6pt}

%\begin{figure}
%	\begin{center}
%		\includegraphics[width=6cm]{example_rho-eps-converted-to.pdf}    % The printed column width is 8.4 cm.
%		\vspace{-9pt}\caption{Typical normalized
%radial growth rates $\rho_{K(\theta)}(\theta),\rho_k(\theta),k\in\mathcal K,\theta\in[-\pi,\pi]$, with {the} same $A_1,A_2,b_1^1,b_2^2$ but different $b_1^2,b_2^1$. (a): $A_ix^*+b_i\not=0,i\in\{1,2\}$ (\emph{Case 1}), (b): $A_ix^*+b_i=0,i\in\{1,2\}$ (\emph{Case 2}). The parameters $b_1^2,b_2^1$ in (a) is obtained by giving a small perturbation on $b_2^1$ and $b_1^2$ in (b).
%\vspace{-24pt}
%}
%		\label{rho}
%
%	\end{center}
%\end{figure}

To discuss how a small perturbation on the parameters in $A_1,A_2$, $b_1,b_2$ affect the stability of the Nash equilibrium $x^*$,
since from (\ref{Nash}) the small perturbations on $a_{11}^1,a_{12}^1,b_1^1,a_{12}^2,a_{22}^2,b_2^2$ change the location of the Nash equilibrium in the state space,
we focus only on the parameters $a_{22}^1,a_{11}^2,b_1^2,b_2^1$ which do not affect the value of $x^*$.
Specifically, consider \emph{Case 2} ($A_ix^*+b_i=0,i\in\{1,2\}$). Then even a small change in $a_{22}^1$ or $b_2^1$ yields $A_1x^*+b_1\not=0$ so that \emph{Case 2} changes to \emph{Case 3} however small the perturbation is. Moreover, if there further exists a small perturbation on $a_{11}^2$ or $b_1^2$, then {$A_2x^*+b_2$ also becomes nonzero} and hence \emph{Case 3} changes to \emph{Case~1}.
For example, it can be seen from Fig.~\ref{rho} that since the small perturbations on $b_1^2,b_2^1$ for $A_ix^*+b_i=0,i\in\{1,2\}$, change the noncooperative system $\mathcal G(J)$ from \emph{Case~2} to \emph{Case~1}, the active mode $K(\theta)$ may drastically change depending on the phase $\theta$ and hence the stability property of the Nash equilibrium $x^*$ may also be affected.

 {It is worth noting that as long as local stability is concerned around a Nash equilibrium $x^*$, the similar results can be drawn for the case of non-quadratic payoff functions which yield nonlinear pseudo-gradient dynamics. Specifically, for a (not necessarily quadratic)~payoff function $J_i(x)$, it can be expressed in the form of
{\setlength\abovedisplayskip{1pt}
\setlength\belowdisplayskip{1pt}
 \begin{eqnarray}\nonumber \label{taler}
\textstyle J_i(x)\!\!\!\!&=&\!\!\!\!\textstyle J_i(x^*)+\big(\frac{\partial J_i(x^*)}{\partial x}\big)^{\mathrm T}\tilde x+\frac{1}{2}{\tilde x}^{\mathrm T}A_i\tilde x+\varepsilon_i(x)\\
\!\!\!\!&=&\!\!\!\!\textstyle \frac{1}{2}x^\mathrm{T}A_ix + b_i^\mathrm{T}x+c_i+\varepsilon_i(x) ,\vspace{-4pt}
\end{eqnarray}
where} $\varepsilon_i(x)$ includes 3rd- or higher-order terms, $A_i\in\mathbb R^{2\times 2}$ is~the Hessian matrix of $J_i(x)$ evaluated at  $x^*$,
 $b_i=\frac{\partial J_i(x^*)}{\partial x}-A_ix^*\in\mathbb R^2$, and $c_i=J_i(x^*)-(\frac{\partial J_i(x^*)}{\partial x})^{\mathrm T}x^*+\frac{1}{2}{ x^*}^{\mathrm T}A_i x^*\!\in\!\mathbb R$.~Noting that $A_i$ in~(\ref{taler}) plays a similar role as the one in (\ref{eqqudractic}), stability analysis around the Nash equilibrium can be similarly conducted as in the theorems and the propositions given in this section.}
\vspace{-12pt}
\section{Illustrative Numerical {Examples}}\label{numerical}\vspace{-2pt}

In this section, we provide {a couple of} numerical examples in order to validate the results {in the paper}.

\begin{figure*}[htbp]
\begin{minipage}{0.31\textwidth}
	\centering	\includegraphics[width=5.05cm]{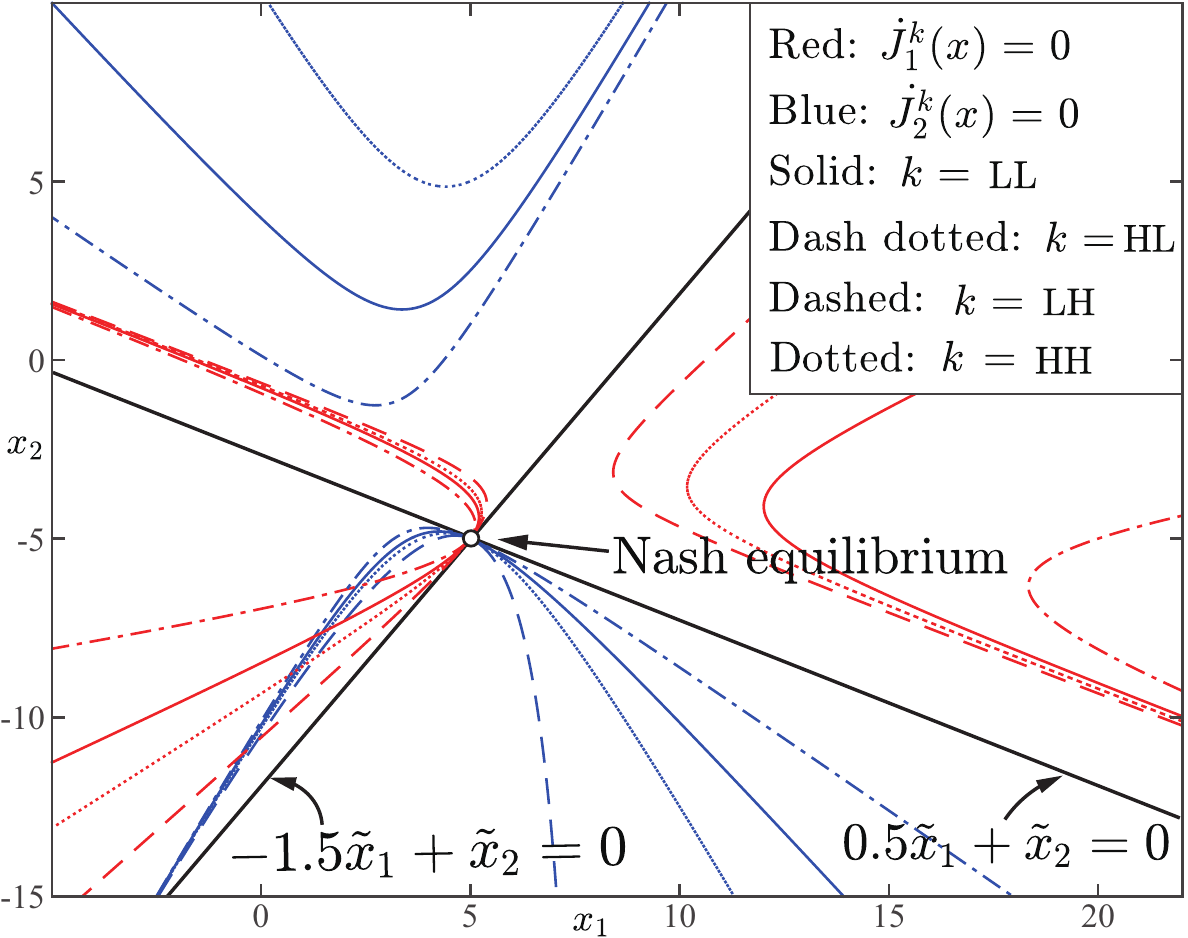}    % The printed column width is 8.4 cm.
	 \vspace{-3pt}	 \caption{The curves of $\dot J^k_i(x)=0$, $i\in\{1,2\}$, $k\in\mathcal K$, in \emph{Example 1}.
}
		\label{example31}
\end{minipage}\hspace{3pt}
\begin{minipage}{0.31\textwidth}
	\centering		\includegraphics[width=5.05cm]{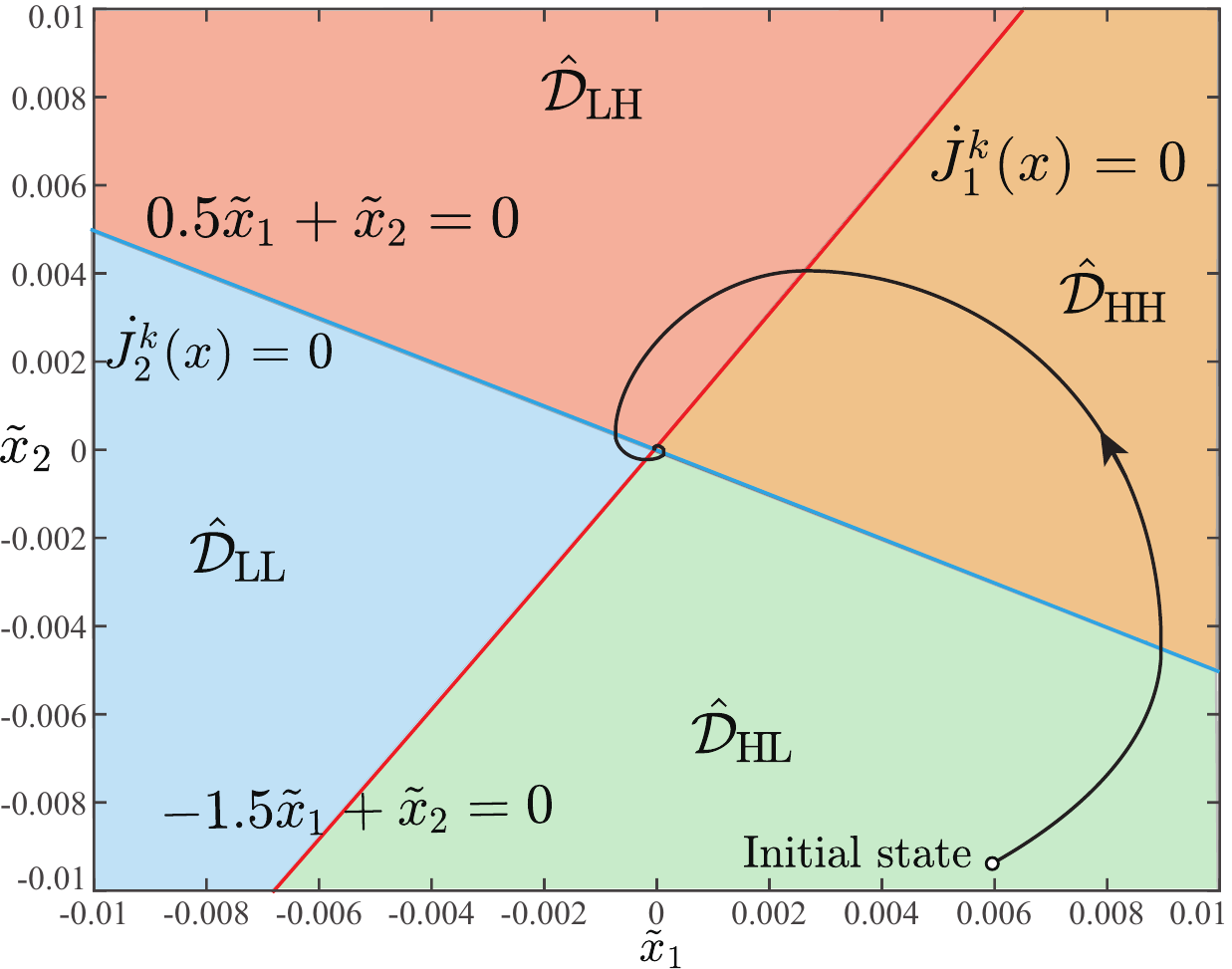}    % The printed column width is 8.4 cm.
 		 \vspace{-3pt} \caption{The approximated domains $\hat{\mathcal D}_k$, $k\in\mathcal K$, and an orbit with $\tilde x=x-x^*$, in \emph{Example 1}.
}
		\label{example34}
\end{minipage}\hspace{-5pt}
\begin{minipage}{0.37\textwidth}
	\centering \includegraphics[width=6.6cm]{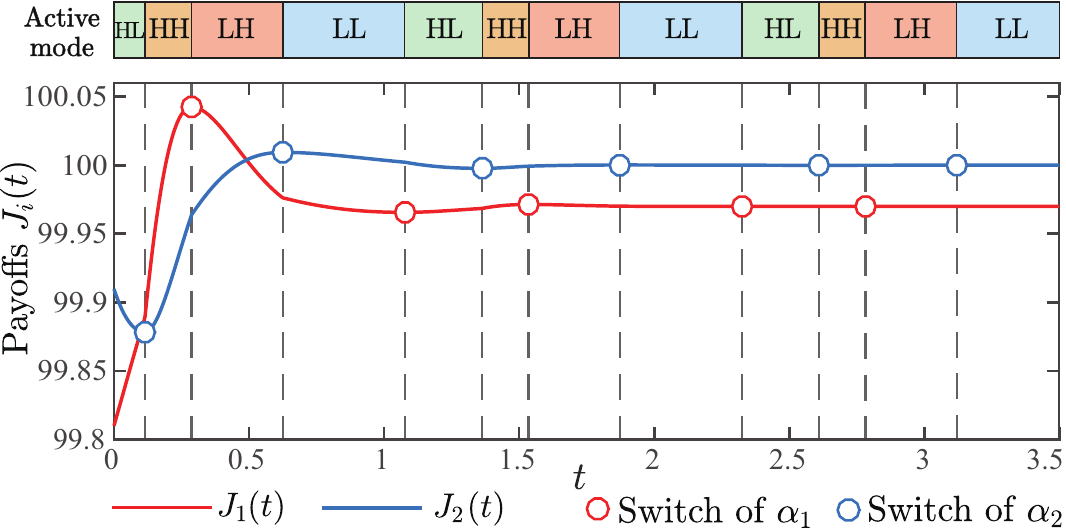}
\caption{Agents' payoffs versus time in \emph{Example 1}.
}
\label{example33}
\end{minipage}\vspace{-10pt}
\end{figure*}

\begin{figure*}[htbp]
\begin{minipage}{0.31\textwidth}
\centering		\includegraphics[width=5.7cm]{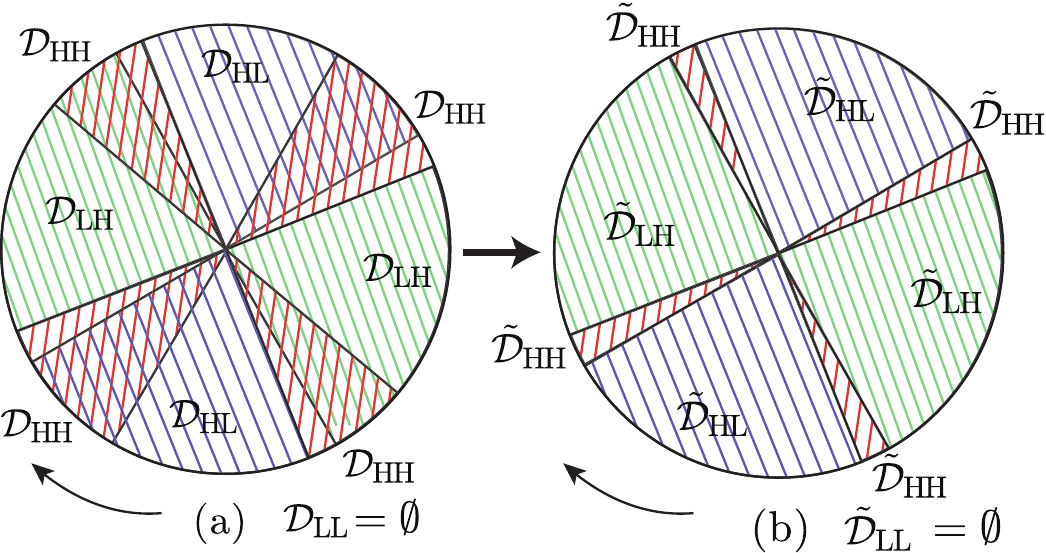}    % The printed column width is 8.4 cm.
	 \caption{The domains of ${\mathcal D}_k$ and $\tilde{\mathcal D}_k$, $k\in\mathcal K$, in \emph{Example 2}. (a): ${\mathcal D}_k$, $k\in\mathcal K$, (b): $\tilde{\mathcal D}_k$, $k\in\mathcal K$ (obtained from (a) with clockwise rotational direction) from which $K(\theta)$ is determined.
}
		\label{example2} 	
\end{minipage}\hspace{3pt}
\begin{minipage}{0.31\textwidth}
\centering		\includegraphics[width=5.03cm]{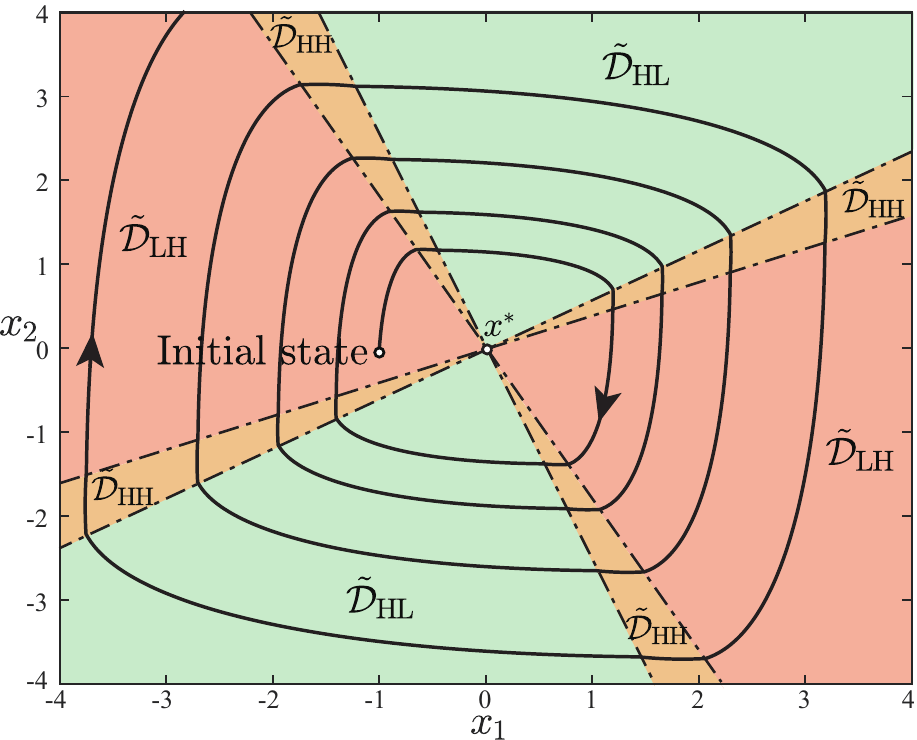}    % The printed column width is 8.4 cm.
		 \vspace{-3pt}\caption{The effective domains $\tilde{\mathcal D}_k$, $k\in\mathcal K$, and an orbit in \emph{Example 2}.}
		\label{example23}	
\end{minipage}\hspace{-3pt}
\begin{minipage}{0.37\textwidth}
%\centering\includegraphics[width=6.7cm]{rho_e2-eps-converted-to.pdf}    % The printed column width is 8.4 cm.
%	\vspace{-21pt}	\caption{Normalized
%radial growth rates $\rho_k(\theta)$, $k\in\mathcal K$, and $\rho_{K(\theta)}(\theta)$, $\theta\in[0,\pi]$, in \emph{Example 2}.}
%		\label{example22}	
\centering 	\includegraphics[width=6.5cm]{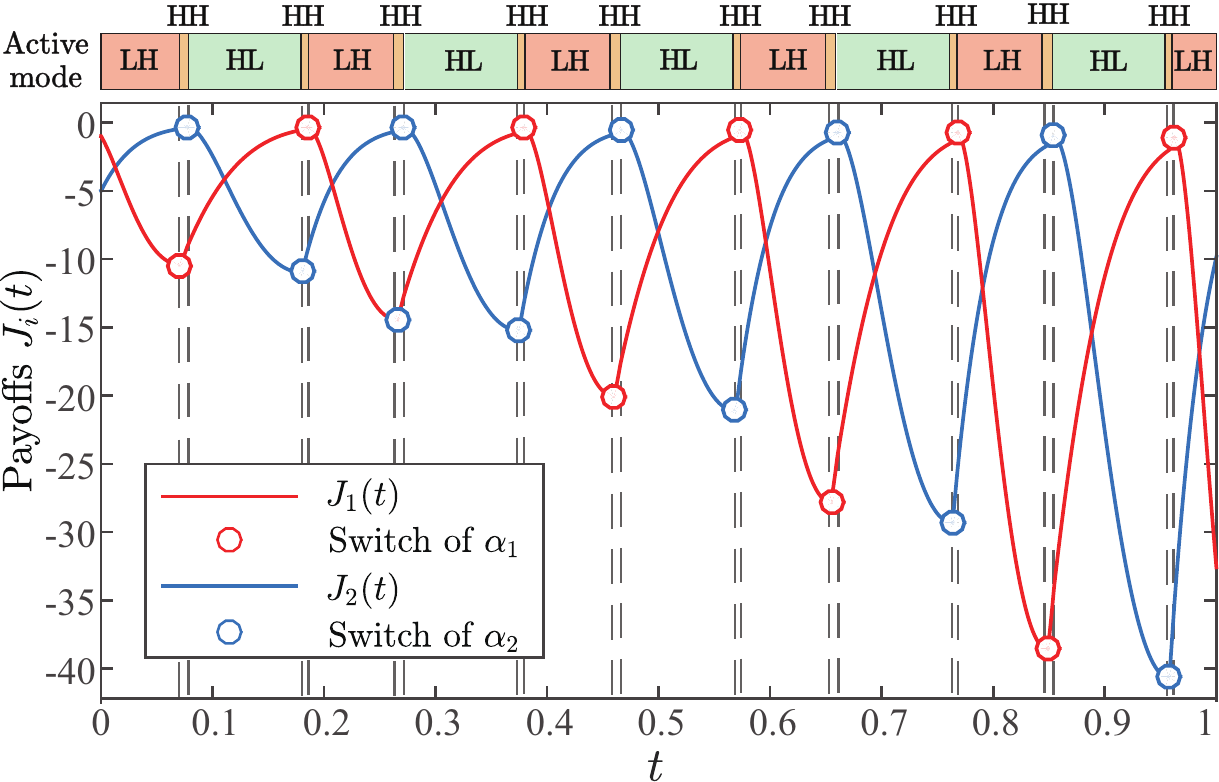}    % The printed column width is 8.4 cm.
		\vspace{-7pt}\caption{Agents' payoffs versus time in \emph{Example~2}.}
		\label{example24}
\end{minipage}\vspace{-15pt}
\end{figure*}

\emph{Example 1.}
Consider the noncooperative system $\mathcal G(J)$ with
$A_1=\Big[\begin{array}{ccc}-2  &  -4\\ -4&  -9 \end{array}\Big]$,
$A_2=\Big[\begin{array}{ccc}-6 & 3\\ 3 & -2\end{array}\Big]$,
$b_1=[-10,-5]^{\mathrm T}$, $ b_2=[30,-25]^{\mathrm T}$, $ c_1=162.47,c_2=0$,
where the Nash equilibrium $x^*=[5,-5]^{\mathrm T}$ satisfies $A_ix^*+b_i\not=0,i\in\{1,2\}$ (\emph{Case~1}).
%\begin{figure}
%	\begin{center}
%		\includegraphics[width=7cm]{J_dot_ex3-eps-converted-to.pdf}    % The printed column width is 8.4 cm.
%		\vspace{-9pt}\caption{The plot of $\dot J^k_i(x)=0$, $i\in\{1,2\}$, $k\in\mathcal K$, in \emph{Example 1}.
%}
%		\label{example31}
%\vspace{-25pt}
%	\end{center}
%\end{figure}
Letting $\alpha_1^{\rm L}=\alpha_2^{\rm L}=1$, $\alpha_1^{\rm H}=2$, $\alpha_2^{\rm H}=3$,~Assumption~\ref{assum4} is satisfied. %the eigenvalues of $\mathbb A_k$, $k=\rm LL,HL,LH,HH$, are respectively given by $-2.0 \pm 3.5i$, $-3.0 \pm 4.8i$, $-4.0 \pm 5.7i$, and $-5.0 \pm 8.4i$.
Figure~\ref{example31} shows the curves of $\dot J^k_i(x)=0,i=$ $1,2$, for all the modes $k\in\mathcal K$.
In this case, %since ${\det(\mathbb A_k^{\mathrm T}+\mathbb A_k)}=15,7,23,95$ ($>0$) for the modes $k={\rm LL,HL,LH,HH}$, respectively,
$\mathbb A_k^{\mathrm T}+\mathbb A_k<0$  for all $k\in\mathcal K$, $\theta \in [0,2 \pi]$, and hence
the normalized radial growth rates $\rho_k(\theta)<0,k\in\mathcal K$, imply $\gamma_{\rm rg}<0$.
%{It can be seen from Fig.~\ref{example35} that the normalized radial growth rate $\rho_{K(\theta)}(\theta)$ less than $0$ for all $\theta\in[0,2\pi)$.}
Hence, it follows from Theorem~\ref{thm2} that the Nash equilibrium $x^*$ is asymptotically stable, which can be verified by the trajectories of states and payoffs shown in Figs.~\ref{example34} and~\ref{example33}.
% {Figure~\ref{example3_f} shows the phase portrait and the payoff functions $J_1(x(t))$ and $J_2(x(t))$ versus time}.

%\begin{figure*}[htbp]
%\begin{minipage}{0.3\textwidth}
%\centering			
%	\includegraphics[width=5.6cm]{rho_e5-eps-converted-to.pdf}    % The printed column width is 8.4 cm.
%		\caption{Normalized
%radial growth rates $\rho_k(\theta)$, $k\in\mathcal K$, and $\rho_{K({\theta})}(\theta)$, $\theta\in[0,2\pi]$, in \emph{Example~3}.
%}
%		\label{example41}
%\end{minipage}\hspace{7pt}
%\begin{minipage}{0.281\textwidth}
%           \includegraphics[width=5cm]{domain_ex5-eps-converted-to.pdf}    % The printed column width is 8.4 cm.
%\caption{The effective domains $\tilde{\mathcal D}_k$, $k\in\mathcal K$, and an orbit in \emph{Example~3}.
%}
%		\label{example42}
%\end{minipage}\hspace{2pt}
%\begin{minipage}{0.37\textwidth}
%\centering \includegraphics[width=6.8cm]{e4_J-eps-converted-to.pdf}    % The printed column width is 8.4 cm.
%		\caption{Agents' payoffs versus time in \emph{Example~3}.
%}
%		\label{example43}
%\end{minipage} \vspace{-15pt}
%\end{figure*}

\emph{Example 2.}
Consider the noncooperative system $\mathcal G(J)$ with
$A_1=\Big[\!\!\begin{array}{ccc}-2  &  4\\ 4&  -10 \end{array}\!\!\Big]^\mathrm{T}$,
$A_2=\Big[\!\!\begin{array}{ccc}-10 & -4\\ -4 & -2\end{array}\!\!\Big]$,
$b_1=b_2=[0,0]^{\mathrm T}$, $ c_1=c_2=0$,
where the Nash equilibrium $x^*=[0,0]^{\mathrm T}$~satisfies $A_ix^*+b_i=0,i\in\{1,2\}$ (\emph{Case~2}).
%\begin{figure}
%	\begin{center}
%\vspace{-5pt}		\includegraphics[width=7.5cm]{e2_J-eps-converted-to.pdf}    % The printed column width is 8.4 cm.
%\vspace{-9pt}
%		\caption{Agents' payoffs versus time in \emph{Example 2}.}
%		\label{example24}
%\vspace{-27pt}
%	\end{center}
%\end{figure}
{Suppose that $\alpha_1^{\rm L}=\alpha_1^{\rm H}=6$, $\alpha_2^{\rm L}=\alpha_2^{\rm H}=9$ for representing the case where the agents are not loss-averse. Then, the eigenvalues of $\mathbb A_{\rm LL}=\mathbb A_{\rm HL}=\mathbb A_{\rm LH}=\mathbb A_{\rm HH}$~are given by $-15.0 \pm 29.2i$, which imply that the Nash equilibrium is stable with the identical subsystem dynamics for all the modes.
Now, suppose that both agents are loss-averse and let} $\alpha_1^{\rm L}=\alpha_2^{\rm L}=1$, then the eigenvalues of $\mathbb A_k$, $k=\rm LL,HL,LH,HH$, are respectively given by $-1.0 \pm 4.0i$, $-7.0 \pm 8.4i$, $-10.0 \pm 8.9i$, and $-15.0 \pm 29.2i$ {so that $\mathbb A_k,k\in\mathcal K$, are still all stable matrices.
Figure~\ref{example2}(a) shows the domains of $\mathcal D_k,k\in\mathcal K$. Noting} that $a_{12}^1>0 $ {and} $ a_{12}^2<0$, it follows from Lemma~\ref{cor1} that the rotational direction is clockwise.
Hence, we re-partition the state space from ${\mathcal D}_k$, $k\in\mathcal K$, {to identify} the effective domains $\tilde{\mathcal D}_k$, $k\in\mathcal K$, as shown in Fig.~\ref{example2}(b), and hence derive the function of $K(\theta)$.
%, and the normalized radial growth {rate} $\rho_{K(\theta)}(\theta)$, $\theta\in[0,\pi]$, is shown in Fig.~\ref{example22}.
Note that the integral of the normalized radial growth rate is
$\gamma_{\rm rg}=\int_0^{2\pi}\rho_{K(\theta)}(\theta){\rm d} \theta=2\int_0^{\pi}\rho_{K(\theta)}(\theta){\rm d} \theta=-0.3224<0. $
Hence, it follows from Theorem~\ref{thm1} that the Nash equilibrium is unstable since $a_{12}^1\gamma_{\rm rg}<0$ {and} $ a_{12}^2\gamma_{\rm rg}>0$ {even though all the subsystem matrices are stable}. The result of Lemma~\ref{cor1} and Theorem~\ref{thm1} can be verified from the trajectories of states and payoff values shown in Figs.~\ref{example23} and \ref{example24}.

\vspace{-8pt}
\section{Conclusion}\label{conclusion}\vspace{-3pt}
We investigated the stability conditions of the noncooperative switched systems with loss-averse agents, where each agent {under pseudo-gradient dynamics exhibits lower sensitivity} for the cases of losing payoffs.
%By involving the loss-aversion-based psychological consideration, we used the lower sensitivity parameter to describe the natural behavior that agents are more cautious to make the decision when they face losing payoff than gaining payoff.
%To characterize the stability property of the Nash equilibrium, we {presented} some general properties of {the system dynamics} for the loss-aversion-based noncooperative switched system in terms of the 4 modes that may arise and the rotational direction of the trajectories.
%We showed that the rotational direction is uniformly the same in the entire state space and explored {how the mode of the switched system transitions}.
%Next, we .
%In addition,
We characterized the notion of the flash switching phenomenon and examined stability properties %of the Nash equilibrium %by using the integral of the normalized radial growth rate
in accordance with the location of the Nash equilibrium for 3 cases.
%As a result,
We revealed %{that the 4 domains corresponding to the 4 modes are invariant with respect to the sensitivity parameter in \emph{Case 1} and}
how the sensitivity parameters   influence the stability property of the system in terms of
the dynamics, partition of the state space, mode transition, and the normalized radial growth rate for each of the 3 cases.
%We provided {several} numerical examples for {illustrating}
%our results. %
 {One of the illustrative examples indicates   that loss-aversion behavior inspired by psychological consideration in prospect theory may result in changing the stability property of the Nash equilibrium from stable to unstable.}
% {.....(Discussion for non-quadratic payoff functions).}
%The extension for the case where the system matrices include real eigenvalues has been considered and submitted for a conference publication.
The future direction may include  {the analysis on the convergence rate and} the extension to higher-dimensional systems  {with more   number of agents and higher dimension of the state. Since the approach with the radial growth rate is  applicable only to planar systems, it is necessary to employ some other methods  in piecewise linear systems theory in higher dimensions.}

\vspace{-8pt}{
\bibliographystyle{IEEEtran}
\bibliography{root}}
%\appendices
 \vspace{-10pt}
\section*{Appendix} \vspace{-3pt}

\emph{Proof of Lemma~\ref{thm_nonzeno}}: %\noindent\textbf{Proof}
First,
by defining
{\setlength\abovedisplayskip{1pt}
\setlength\belowdisplayskip{1pt}\begin{eqnarray}\label{delta_11}
\Delta_1^1(x)\!\!\!&\triangleq&\!\!\!(a_{11}^1x_1+a_{12}^1x_2+b_1^1)^2\geq0,\\ \label{delta_12}
\Delta_1^2(x)\!\!\!&\triangleq&\!\!\!(a_{12}^1x_1+a_{22}^1x_2+b_2^1)(a_{12}^2x_1+a_{22}^2x_2+b_2^2),\quad\,\\
\label{delta_21}
\Delta_2^1(x)\!\!\!&\triangleq&\!\!\!(a_{11}^2x_1+a_{12}^2x_2+b_1^2)(a_{11}^1x_1+a_{12}^1x_2+b_1^1),\\ \label{delta_22}
\Delta_2^2(x)\!\!\!&\triangleq&\!\!\!(a_{12}^2x_1+a_{22}^2x_2+b_2^2)^2\geq0,
\end{eqnarray}
the} functions in (\ref{J_ik}) can be   {calculated with (\ref{Nash}) as}
{\setlength\abovedisplayskip{2pt}
\setlength\belowdisplayskip{2pt}\begin{eqnarray}\label{J_dot_sum}
\dot J_i^k(x)=\alpha^k_1\Delta_i^1(x)+\alpha^k_2\Delta_i^2(x), \quad i\in\{1,2\}, \quad k\in\mathcal K.
\end{eqnarray}
Let $\delta_i\triangleq\alpha_i^{\rm H}-\alpha_i^{\rm L}\geq0$, $i=1,2$.
%{\setlength\abovedisplayskip{1pt}
%\setlength\belowdisplayskip{1pt}\begin{eqnarray}\label{J_dot_begin}
%\dot J^{\rm LL}_1(x)\!\!\!\!&=&\!\!\!\!\alpha_1^{\rm L}\Delta_1^1(x)+\alpha_2^{\rm L}\Delta_1^2(x),\\
%\dot J^{\rm LL}_2(x)\!\!\!\!&=&\!\!\!\!\alpha_1^{\rm L}\Delta_2^1(x)+ \alpha_2^{\rm L}\Delta_2^2(x),\\
%\dot J^{\rm HL}_1(x)\!\!\!\!&=&\!\!\!\!\alpha_1^{\rm H}\Delta_1^1(x)+\alpha_2^{\rm L}\Delta_1^2(x)=\dot J^{\rm LL}_1(x)+\delta_1\Delta_1^1(x),\quad\\
%\dot J^{\rm HL}_2(x)\!\!\!\!&=&\!\!\!\!\alpha_1^{\rm H}\Delta_2^1(x)+ \alpha_2^{\rm L}\Delta_2^2(x)=\dot J^{\rm LL}_2(x)+\delta_1\Delta_2^1(x),\\
%\dot J^{\rm LH}_1(x)\!\!\!\!&=&\!\!\!\!\alpha_1^{\rm L}\Delta_1^1(x)+\alpha_2^{\rm H}\Delta_1^2(x)=\dot J^{\rm LL}_1(x)+\delta_2\Delta_1^2(x),\quad\\
%\label{J_dot3_end}
%\dot J^{\rm LH}_2(x)\!\!\!\!&=&\!\!\!\!\alpha_1^{\rm L}\Delta_2^1(x)+ \alpha_2^{\rm H}\Delta_2^2(x)=\dot J^{\rm LL}_2(x)+\delta_2\Delta_2^2(x),\quad\\
%\nonumber
%\dot J^{\rm HH}_1(x)\!\!\!\!&=&\!\!\!\!\alpha_1^{\rm H}\Delta_1^1(x)+\alpha_2^{\rm H}\Delta_1^2(x)\\
%                     &=&\!\!\!\!\dot J^{\rm LL}_1(x)+\delta_1\Delta_1^1(x)+\delta_2\Delta_1^2(x),\\ \nonumber
%\dot J^{\rm HH}_2(x)\!\!\!\!&=&\!\!\!\!\alpha_1^{\rm H}\Delta_2^1(x)+ \alpha_2^{\rm H}\Delta_2^2(x)\\ \label{J_dot_end}
%&=&\!\!\!\!\dot J^{\rm LL}_2(x)+\delta_1\Delta_2^1+\delta_2\Delta_2^2.
%\end{eqnarray}
Now, we suppose ${\mathcal D}_{\rm LL}\not=\mathbb R^2$ so that there exists $\bar x\in\mathbb R^2$ such that $\bar x\not\in {\mathcal D}_{\rm LL}$. In this case, there are three cases in terms of $\bar x$ that may happen: $\dot J^{\rm LL}_1(\bar x)>0\wedge\dot J^{\rm LL}_2(\bar x)\leq 0$; $\dot J^{\rm LL}_1(\bar x)\leq0\wedge \dot J^{\rm LL}_2(\bar x)>0$; $\dot J^{\rm LL}_1(\bar x)>0\wedge\dot J^{\rm LL}_2(\bar x)>0$.
For the case of $\dot J^{\rm LL}_1(\bar x)>0\wedge\dot J^{\rm LL}_2(\bar x)\leq 0$, since
$\Delta_1^1(\bar x)\geq0$, we have $\dot J^{\rm HL}_1(\bar x)=\dot J^{\rm LL}_1(\bar x)+\delta_1\Delta_1^1(\bar x)>0$. Moreover, since $\dot J^{\rm LL}_2(\bar x)=\alpha_1^{\rm L}\Delta_2^1(x)+ \alpha_2^{\rm L}\Delta_2^2(x)\leq0$ and $\Delta_2^2(\bar x)\geq0$ imply $\Delta_2^1(\bar x)\leq0$, we have $\dot J^{\rm HL}_2(\bar x)=\dot J^{\rm LL}_2(\bar x)+\delta_1\Delta_2^1(\bar x)\leq0$. Hence, $\bar x\in {\mathcal D}_{\rm HL}$.
  {The other cases can be similarly handled with the conclusion of  $\bar x\in {\mathcal D}_{\rm LH}$ or $\bar x\in {\mathcal D}_{\rm HH}$ (see \cite{yan2022stability} for the complete proof).}
%Case 2 ($\dot J^{\rm LL}_1(\bar x)\leq0\wedge \dot J^{\rm LL}_2(\bar x)>0$): In this case, since
%$\dot J^{\rm LL}_1(\bar x)=\alpha_1^{\rm L}\Delta_1^1(x)+\alpha_2^{\rm L}\Delta_1^2(x)\leq0$ and $\Delta_1^1(\bar x)\geq0$ imply $\Delta_1^2(\bar x)\leq0$, we have $\dot J^{\rm LH}_1(\bar x)=\dot J^{\rm LL}_1(\bar x)+\delta_2\Delta_1^2(\bar x)\leq0$.
%Moreover, since $\Delta_2^2(\bar x)\geq0$, we have
%$\dot J^{\rm LH}_2(\bar x)=\dot J^{\rm LL}_2(\bar x)+\delta_2\Delta_2^2(\bar x)>0$. Hence, $\bar x\in {\mathcal D}_{\rm LH}$.
%
%Case 3 ($\dot J^{\rm LL}_1(\bar x)>0\wedge\dot J^{\rm LL}_2(\bar x)>0$): In this case, note that since $\Delta_1^1(\bar x)\geq0$ and $\Delta_2^2(\bar x)\geq0$, the inequalities $\dot J^{\rm HL}_1(\bar x)=\dot J^{\rm LL}_1(\bar x)+\delta_1\Delta_1^1(\bar x)>0$ and $\dot J^{\rm LH}_2(\bar x)=\dot J^{\rm LL}_2(\bar x)+\delta_2\Delta_2^2(\bar x)>0$ must hold.
%Now, we further suppose that $\bar x\not\in {\mathcal D}_{\rm HL}$ and $\bar x\not\in {\mathcal D}_{\rm LH}$ hold, i.e., we suppose that $\dot J^{\rm HL}_2(\bar x)>0\wedge\dot J^{\rm LH}_1(\bar x)>0$ holds.
%Then, since $\dot J^{\rm HH}_1(\bar x)=\dot J^{\rm LH}_1(\bar x)+\delta_1\Delta_1^1(\bar x)>0\wedge\dot J^{\rm HH}_2(\bar x)=\dot J^{\rm HL}_2(\bar x)+\delta_2\Delta_2^2(\bar x)>0$, we have $\bar x\in {\mathcal D}_{\rm HH}$.
Thus, for any $\bar x\in\mathbb R^2$, there exist $k\in\mathcal K$ s.t. $\bar x\in\mathcal D_k$.%, which completes the proof.
$\hfill\square$

\emph{Proof of Lemma~\ref{cor1}}: %\noindent\textbf{Proof}
Note that $a_{12}^1<0\wedge a_{12}^2>0$ and $a_{12}^1>0\wedge a_{12}^2<0$ imply
 that the diagonal elements of $P_k$ %satisfy $\alpha_2^ka_{12}^2>0\wedge-\alpha_1^ka_{12}^1>0$, $k\in\mathcal K$, and $\alpha_2^ka_{12}^2<0\wedge-\alpha_1^ka_{12}^1<0$, $k\in\mathcal K$,
 are all positive and negative, respectively, and hence $P_k>0$ (resp., $P_k<0$), $k\in\mathcal K$, because {Assumption~\ref{assum4} implies $\det P_k>0$}. Thus, the result is immediate since
$\dot\theta_k=\eta^{\mathrm T}(\theta)P_k\eta(\theta)$. %(resp., $<0$) for $a_{12}^1<0\wedge a_{12}^2>0$ (resp., $a_{12}^1>0\wedge a_{12}^2<0$), $\theta\in[0,2\pi)$ and $k\in\mathcal K$.
$\hfill\square$

\noindent\emph{Proof of Proposition~\ref{slope}}: %\noindent\textbf{Proof}
  {The proof is immediate by checking
the values of $\frac{\partial \dot J_i^k(x^*)}{\partial x_j}$, $i,j=1,2$,  $k\in \mathcal K$ (see \cite{yan2022stability} for the complete proof).~$\hfill\square$}

\emph{Proof of Lemma~\ref{lemma_4}}:
  {As the curve $\dot J_1^k(x)=0$ (resp., $\dot J_2^k(x)=0$) is linearized by the straight line (\ref{approximated}) (resp., (\ref{approximated2})) for all $k\in\mathcal K$ (Proposition~\ref{slope}), the proof is immediate by checking whether the 4 domains $\{x\in\mathbb R^2:\dot J^k_i( x)\geq0\}$, $k\in\mathcal K$, share exactly the same half plane in the neighborhood of $x^*$, which  is proved by  the fact that
 % so that the sign of $\dot J^k_i(\hat x)$ is indicating the sign of $\dot J^k_i(x)$ in the left side (resp., right side) with $\varepsilon<0$ (resp., $\varepsilon>0$) in the neighborhood of $x^*$.
 {\setlength\abovedisplayskip{1pt}
\setlength\belowdisplayskip{1pt}
\begin{align}\nonumber
\textstyle\dot J^k_1(\hat x) %=&\textstyle(\alpha^k_1a_{11}^1a_{11}^1+\alpha^k_2a_{12}^1a_{12}^2)\varepsilon^2\\ \nonumber \label{sign_right1}
%&\textstyle+\varepsilon \alpha^k_2a_{12}^2(a_{12}^1 x_1^*+a_{22}^1x_2^*+b_2^1)\\
\thickapprox \varepsilon \alpha_2^ka_{12}^2(a_{12}^1 x_1^*+a_{22}^1x_2^*+b_2^1),\quad k\in\mathcal K,\\ \nonumber
\textstyle\dot J^k_2(\hat x) %=&\textstyle(\alpha^k_2a_{12}^2a_{12}^2+\alpha^k_1a_{11}^1a_{11}^2)\varepsilon^2\\ \nonumber
%&\textstyle+\varepsilon\alpha^k_1a_{11}^1(a_{11}^2 x_1^*+a_{12}^2x_2^*+b_1^2)\\ \label{sign_right2}
\thickapprox \varepsilon \alpha^k_1a_{11}^1(a_{11}^2 x_1^*+a_{12}^2x_2^*+b_1^2),\quad k\in\mathcal K,
\end{align}
hold for }
%&=&\!\!\!\!\alpha_1(a_{11}^1\varepsilon)^2+\alpha_2a_{12}^2(a_{12}^1\hat x_2+a_{22}^1x_1^*+b_2^1)\\
%&=&\!\!\!\!(\alpha^k_2a_{12}^2a_{12}^2+\alpha^k_1a_{11}^1a_{11}^2)\varepsilon^2\\ \nonumber
%&&\!\!\!\!+\varepsilon\alpha^k_1a_{11}^1(a_{11}^2 x_1^*+a_{12}^2x_2^*+b_1^2)\\
%\label{sign_right2}
$\hat x\triangleq[x_1^*+\varepsilon,x_2^*]^{\mathrm T}$ with an infinitesimal number $\varepsilon$ (see \cite{yan2022stability} for the complete proof).}
$\hfill\square$

\emph{Proof of Theorem~\ref{thm2}}: %\noindent\textbf{Proof}
First, note that $
\gamma_{\rm rg}=\int_{\theta_0}^{\theta_0+2\pi}\!\!\!\!\rho_{K(\theta)}(\theta){\rm d} \theta$ $=\int_{\theta_0}^{\theta_0+2\pi}\!\!\frac{1}{r}\frac{{\rm d}r}{{\rm d}\theta}{\rm d} \theta=\log\frac{r_{\theta_0+2\pi}}{r_{\theta_0}},$
 where $\frac{r_{\theta_0+2\pi}}{r_{\theta_0}}$ represents the ratio of the distances between the Nash equilibrium $x^*$  and the states when the state travels for one round from the phase $\theta_0$ to $\theta_0+2\pi$. For the counterclockwise case (i.e., $a_{12}^1<0\wedge a_{12}^2>0$), $\gamma_{\rm rg}<0$ (resp., $\gamma_{\rm rg}>0$) implies that the state is coming closer to (resp., farther from) $x^*$ under (\ref{PLS}) after it travels for one round. For the clockwise case (i.e., $a_{12}^1>0\wedge a_{12}^2<0$), the opposite is true. Hence, if $a_{12}^1\gamma>0\wedge a_{12}^2\gamma<0$ (resp., $a_{12}^1\gamma<0\wedge a_{12}^2\gamma>0$), then noting $a_{12}^1a_{12}^2<0$, the Nash equilibrium $x^*$ is asymptotically stable (resp., unstable).
% if $a_{12}^1\gamma_{\rm rg}>0\wedge a_{12}^2\gamma_{\rm rg}<0$ (resp., $a_{12}^1\gamma_{\rm rg}<0\wedge a_{12}^2\gamma_{\rm rg}>0$), then noting $a_{12}^1a_{12}^2<0$, the Nash equilibrium $x^*$ is asymptotically stable (resp., unstable).
$\hfill\square$

\emph{Proof of Theorem~\ref{thm1}}: %\noindent\textbf{Proof}
The proof for \emph{1)} and \emph{3)} is similar to the proof of Theorem~\ref{thm2}. For both cases of counterclockwise and clockwise trajectories, $\gamma_{\rm rg}=0$ implies that the trajectory goes back to the same point when it travels for one round from the phase $\theta_0$ to $\theta_{0}+2\pi$.
Now, \emph{2)} is immediate.
$\hfill\square$

\emph{Proof of Theorem~\ref{transition_thm}}: %   \noindent\textbf{Proof}
First, we prove $K(\theta(t_2))=K(\theta(t_2^+))\in\{\rm LH,HL\}$ (implying that $t_2$ is not a flash switching instant). %and the active mode must be either $\rm HL$ or $\rm LH$ for $t=t_2$.
To this end,
%note that the condition of $K(\theta(t))=\rm LL$ or $\rm HH$ for $t_1< t<t_2$ implies that the domain $\tilde {\mathcal D}_{\rm LL}\subseteq {\mathcal D}_{\rm LL}$ or $\tilde {\mathcal D}_{\rm HH}\subseteq {\mathcal D}_{\rm HH}$ exists.
let the state at $t_2$ as $\bar x$  {and characterize cases} %Note that under Assumption~\ref{assum5}, $\bar x$ satisfies
 {in terms~of $K(\theta(t_2^-))$ and $\bar x$. %there are only 2 cases given by $K(\theta(t_2^-))=\rm LL$ and $K(\theta(t_2^-))=\rm HH$.
%For each case, it can be further separated to subcases where agent 1 or agent 2 switches its sensitivity at $t_2$.
For example, consider $K(\theta(t_2^-))=\rm LL$ (i.e., $K(\theta(t))=\rm LL$, $t_1<t<t_2$) and  $\dot J^{\rm LL}_1(\bar x)=0\wedge\dot J^{\rm LL}_2(\bar x)<0$ (i.e., $K(\theta(t_2))=\rm HL$). In this case,}
Since $\Delta_2^2(\bar x)\geq0$ and $\dot J^{\rm LL}_2(\bar x)=\alpha_1^{\rm L}\Delta_2^1(\bar x)+ \alpha_2^{\rm L}\Delta_2^2(\bar x)<0$ imply $\Delta_2^1(\bar x)<0$, we have $\dot J^{\rm HL}_2( t_2^+ )\approx\dot J^{\rm HL}_2(\bar x)=\dot J^{\rm LL}_2(\bar x)+\delta_1\Delta_2^1(\bar x)<0$.
Note that when $\Delta_1^1(\bar x)>0$, we have $\dot J^{\rm HL}_1(t_2^+)\approx\dot J^{\rm HL}_1(\bar x)=\dot J^{\rm LL}_1(\bar x)+\delta_1\Delta_1^1(\bar x)>0$.
%Hence, $x(t_2^+)=\bar x\in {\rm int}\, {\mathcal D}_{\rm HL}$.
Alternatively, when $\Delta_1^1(\bar x)=\dot J^{\rm LL}_1(\bar x)=0$, we have $\Delta_1^2(\bar x)=\dot J^{\rm HL}_1(\bar x)=0$.
By neglecting the second-order infinitesimal in $\Delta_1^1(x(t_2^+))$ in (\ref{J_dot_sum}),
it follows that
$\dot J^{k}_1(t_2^+)\approx \alpha_2^{k}\Delta_1^2(x(t_2^+))$ holds for $k\in\mathcal K$.
Hence, since $\dot J^{\rm LL}_1(t_2^+)>0$,
it follows that
$\dot J^{\rm HL}_1(t_2^+)\approx \alpha_2^{\rm L}\Delta_1^2(x(t_2^+))\approx \dot J^{\rm LL}_1(t_2^+)>0$.
Consequently, $x(t_2^+)\in {\rm int}\, {\mathcal D}_{\rm HL}$ holds for both the two cases above in terms of $\Delta_1^1(\bar x)$ and hence
%agents hold the mode $\rm HL$ after the switching, i.e.,
$K(\theta(t_2^+))=K(\theta(t_2))=\rm HL$ holds.
%1-2) Consider the case $\dot J^{\rm LL}_2(\bar x)=0\wedge\dot J^{\rm LL}_1(\bar x)<0$ where agent 2 switches its sensitivity at $t_2$ from $\alpha_2^{\rm L}$ to $\alpha_2^{\rm H}$ and hence $K(\theta(t_2))=\rm LH$.
%Since $\Delta_1^1(\bar x)\geq0$ and
%$\dot J^{\rm LL}_1(\bar x)=\alpha_1^{\rm L}\Delta_1^1(\bar x)+\alpha_2^{\rm L}\Delta_1^2(\bar x)<0$ imply $\Delta_1^2(\bar x)<0$, we have $\dot J^{\rm LH}_1(x(t_2^+))\approx\dot J^{\rm LH}_1(\bar x)=\dot J^{\rm LL}_1(\bar x)+\delta_2\Delta_1^2(\bar x)<0$.
%Note that when $\Delta_2^2(\bar x)>0$, we have
%$\dot J^{\rm LH}_2(x(t_2^+))\approx\dot J^{\rm LH}_2(\bar x)=\dot J^{\rm LL}_2(\bar x)+\delta_2\Delta_2^2(\bar x)>0$.
%Alternatively, when $\Delta_2^2(\bar x)=\dot J^{\rm LL}_2(\bar x)=0$, we have $\Delta_2^1(\bar x)=\dot J^{\rm LH}_2(\bar x)=0$.
%Hence, since $\dot J^{\rm LL}_1(x(t_2^+))>0$,
%it follows from (\ref{tool22}) that
%$\dot J^{\rm LH}_2(x(t_2^+))\approx \alpha_1^{\rm L}\Delta_2^1(x(t_2^+))\approx\dot J^{\rm LL}_2(x(t_2^+))>0$.
%Consequently, $x(t_2^+)\in {\rm int}\, {\mathcal D}_{\rm LH}$ holds for both the two cases above in terms of $\Delta_2^2(\bar x)$ and hence
%%agents hold the mode $\rm LH$ after the switching, i.e.,
%$K(\theta(t_2^+))=K(\theta(t_2))=\rm LH$ holds.
%Case 2: $K(\theta(t_2^-))=\rm HH$ (i.e., $K(\theta(t))=\rm HH$, $t_1<t<t_2$).
 {The proof for the other cases can be similarly handled with the conclusion of $K(\theta(t_2))=K(\theta(t_2^+))\in\{\rm LH,HL\}$ (see \cite{yan2022stability} for the complete proof).}
Thus,
 $K(\theta(t_2^+))=K(\theta(t_2))\in\{\rm LH, HL\}$ for $K(\theta(t_2^-))\in\{\rm LL, HH\}$. % and hence $t_2$ is not a flash switching instant.
Furthermore, since $K(\theta(t_1^+))\in\{\rm LL, HH\}$, it follows that $K(\theta(t_1^-))\not\in\{\rm LL, HH\}$, i.e., $K(\theta(t_1^-))\in\{\rm LH, HL\}$,
%.In addition,
%$K(\theta(t_1))\not=K(\theta(t_1^+))$ does not hold since $K(\theta(t_1)),K(\theta(t_1^+))\in\{\rm LL, HH\}$.
which implies that $t_1$ is not a flash switching instant, either.

Next, we prove $K(\theta(t_1^-))\not=K(\theta(t_2^+))$ for sign-indefinite $A_1,A_2$.
To this end, we show that
$\mathcal D_{\rm LH}$, $\mathcal D_{\rm HL}$ are never composed of 4 convex cones.
%Without loss of generality, we give the proof for $k={\rm LH}$.
Suppose, \emph{ad absurdum}, $\mathcal D_{\rm LH}=\{ x\in\mathbb R^2:\dot J^{\rm LH}_1( x)\leq0\}\cap\{ x\in\mathbb R^2:\dot J^{\rm LH}_2(x)\geq0\}$ is composed of 4 convex cones in $\mathcal G(J)$
with a certain set of sensitivity parameters $(\alpha_1^{\rm H},\alpha_1^{\rm L},\alpha_2^{\rm H},\alpha_2^{\rm L})=(\tilde\alpha_1^{\rm H},\tilde \alpha_1^{\rm L},\tilde\alpha_2^{\rm H},\tilde\alpha_2^{\rm L})$.
In this case, from a geometric consideration of the domains, it follows that
$
\{ x\in\mathbb R^2:\dot J^{\rm LH}_1( x)\leq0\}\cup\{ x\in\mathbb R^2:\dot J^{\rm LH}_2(x)\geq0\}=\mathbb R^2.
$
Next, consider a set of sensitivity parameters $(\alpha_1^{\rm H},\alpha_1^{\rm L},\alpha_2^{\rm H},\alpha_2^{\rm L})$ with ${\alpha}_1^{\rm H}=\tilde\alpha_1^{\rm L}$ and ${\alpha}_2^{\rm L}=\tilde\alpha_2^{\rm H}$.
Note that since the sensitivity profile $ {\alpha}^{\rm HL}={\rm diag}[\tilde\alpha_1^{\rm L},\tilde\alpha_2^{\rm H}]$ for the second set is the same as the value of ${\alpha}^{\rm LH}={\rm diag}[\tilde\alpha_1^{\rm L},\tilde\alpha_2^{\rm H}]$ in the first set,
the domain {\setlength\abovedisplayskip{1pt}
\setlength\belowdisplayskip{1pt}
\begin{align}\nonumber
\textstyle {\rm int\,}\mathcal D_{\rm HL}\!&=\!\textstyle\{ x\in\mathbb R^2:\dot { J}^{\rm HL}_1( x)>0\}\cap\{ x\in\mathbb R^2:\dot { J}^{\rm HL}_2(x)<0\} \\ \nonumber
\!&=\!\textstyle\{ x\in\mathbb R^2:\dot J^{\rm LH}_1( x)>0\}\cap\{ x\in\mathbb R^2:\dot J^{\rm LH}_2(x)<0\}\\ \nonumber
\!&=\!\textstyle(\mathbb R^2\!\setminus\!\{ x\in\mathbb R^2\!:\!\dot J^{\rm LH}_1( x)\!\leq0\})\!\cap\!\{ x\in\mathbb R^2\!:\!\dot J^{\rm LH}_2(x)\!<0\}\\ \nonumber
\!&=\!\textstyle(\{ x\!\in\!\mathbb R^2\!:\!\dot J^{\rm LH}_2( x)\geq0\}\!\setminus\!\mathcal D_{\rm LH})\cap\!\{ x\!\in\!\mathbb R^2\!:\!\dot J^{\rm LH}_2(x)\!<0\}
\end{align}
  {is empty, which contradicts with Lemma 2 in \cite{yan2022stability}.}}  % in that ${\rm int}\,\mathcal D_{\rm HL}$ is non-empty for any sign-indefinite $A_1$, $A_2$ and $\alpha_1^{\rm H},\alpha_1^{\rm L},\alpha_2^{\rm H},\alpha_2^{\rm L}\in\mathbb R_+$.
Thus, $\mathcal D_{\rm LH}$ is never composed of 4 convex cones. The proof for $\mathcal D_{\rm HL}$ can be similarly handled.
%The proof for the case of $k=\rm HL$ can be similarly handled
  {Now,} suppose, \emph{ad absurdum}, that $K(\theta(t_1^-))=K(\theta(t_2^+))\in\{\rm LH,HL\}$. Since the rotational direction of the trajectories are consistently the same in $\mathbb R^2$,
it follows that
%$K(\theta(t_1^-))=K(\theta(t_2^+))\in\{\rm LH,HL\}$ holds for , only if
$\tilde{ \mathcal D}_{\rm LL}\cup \tilde{ \mathcal D}_{\rm LH}=\mathbb R^2$ or $\tilde{ \mathcal D}_{\rm LL} \cup \tilde{ \mathcal D}_{\rm HL}=\mathbb R^2$
must hold for $K(\theta(t))=\rm LL$, $t_1< t<t_2$,   {which also contradicts with Lemma 2 in \cite{yan2022stability}}. % implies that $\tilde{ \mathcal D}_{\rm LL}\cup \tilde{ \mathcal D}_{\rm LH}\not=\mathbb R^2$, $\tilde{ \mathcal D}_{\rm LL} \cup \tilde{ \mathcal D}_{\rm HL}\not=\mathbb R^2$.
%$K(\theta(t_1^-))\not=K(\theta(t_2^+))$, i.e.,
(The case of $K(\theta(t))=\rm HH$, $t_1<t<t_2$, can be similarly handled.)
%Consequently,
%$(K(\theta(t_1^-)),K(\theta(t_2^+)))\in\{(\rm LH,HL),(HL,LH)\}$ for sign-indefinite $A_1$, $A_2$.
Thus, the proof is complete.
 $\hfill\square$

\emph{Proof of Theorem~\ref{thm3}}: %\noindent\textbf{Proof}
The proof is similar to the proof of Theorems~\ref{thm2} and \ref{thm1}.
$\hfill\square$

\emph{Proof of Proposition~\ref{prop2}}: % \noindent\textbf{Proof}
 {Let $t_1,t_2$ be two consecutive switching instants. The result follows from  the fact that if  $K(\theta(t))=\rm LL$ or $\rm HH$ (resp., $\rm LH$ or $\rm HL$) for $t_1<t<t_2$, then $K(\theta(t_1^-))$, $K(\theta(t_2^+))\in\{\rm LH, HL\}$ (resp., $\{\rm LL, HH\}$) \cite{yan2022stability}.}
$\hfill\square$

\emph{Proof of Proposition~\ref{no_jump}}: %\noindent\textbf{Proof}
The proof is immediate by checking the values of $\rho_k(\theta)$ at the {specified} phases.
$\hfill\square$
\end{document}